\documentclass[prodmode,acmtkdd]{acmsmall} 

\usepackage[ruled]{algorithm2e}

\SetAlFnt{\small}
\SetAlCapFnt{\small}
\SetAlCapNameFnt{\small}
\SetAlCapHSkip{0pt}
\IncMargin{-\parindent}
\usepackage{makecell}
\usepackage{multirow}
\usepackage{pbox}
\usepackage{subfigure}
\usepackage{times}
\usepackage{tabu}
\usepackage{tabularx}
\usepackage{verbatim}
\usepackage{url}
\usepackage{amsmath}

\usepackage{amsthm}
\usepackage{amssymb,bm}
\usepackage{float}
\usepackage{footmisc}
\usepackage{lipsum}
\usepackage{booktabs}
\usepackage{breqn}
\usepackage{graphicx,epstopdf}
\usepackage{tikz}
\usetikzlibrary{automata,arrows}
\newcommand{\norm}[1]{\ensuremath{\lVert{#1}\rVert}}
\DeclareMathOperator*{\argmin}{arg\,min}
\theoremstyle{problem}
\newtheorem{problem}{Problem}

\acmVolume{12}
\acmNumber{2}
\acmArticle{23}
\acmYear{2018}
\acmMonth{3}

\setcopyright{acmcopyright}

\doi{10.1145/3127873}

\issn{1556-4681}

\begin{document}

\markboth{G. Hu et al.}{Collaborative Filtering with Topic and Social Latent Factors Incorporating Implicit Feedback}

\title{Collaborative Filtering with Topic and Social Latent Factors Incorporating Implicit Feedback}
\author{Guang-Neng Hu
\affil{Nanjing Univeristy}
Xin-Yu Dai*
\affil{Nanjing Univeristy}
Feng-Yu Qiu
\affil{Nanjing Univeristy}
Rui Xia
\affil{Nanjing University of Science and Technology}
Tao Li
\affil{Nanjing University of Posts and Telecommunications, Florida International University}
Shu-Jian Huang
\affil{Nanjing Univeristy}
Jia-Jun Chen
\affil{Nanjing Univeristy}
}

\begin{abstract}
Recommender systems (RSs) provide an effective way of alleviating the information overload problem by selecting personalized items for different users. Latent factors based collaborative filtering (CF) has become the popular approaches for RSs due to its accuracy and scalability. Recently, online social networks and user-generated content provide diverse sources for recommendation beyond ratings. Although  {\em social matrix factorization} (Social MF) and {\em topic matrix factorization} (Topic MF) successfully exploit social relations and item reviews, respectively; both of them ignore some useful information. In this paper, we investigate the effective data fusion by combining the aforementioned approaches. First, we propose a novel model {\em \mbox{MR3}} to jointly model three sources of information (i.e., ratings, item reviews, and social relations) effectively for rating prediction by aligning the latent factors and hidden topics. Second, we incorporate the implicit feedback from ratings into the proposed model to enhance its capability and to demonstrate its flexibility. We achieve more accurate rating prediction on real-life datasets over various state-of-the-art methods. Furthermore, we measure the contribution from each of the three data sources and the impact of implicit feedback from ratings, followed by the sensitivity analysis of hyperparameters. Empirical studies demonstrate the effectiveness and efficacy of our proposed model and its extension.
\end{abstract}

%
%

\ccsdesc[500]{Information systems~Recommender systems}
\ccsdesc[500]{Human-centered computing~Collaborative filtering}

%
%


\keywords{Recommendation Systems, Collaborative Filtering, Implicit Feedback, Hidden Topics, Latent Social Factors}

\acmformat{Guang-Neng Hu, Xin-Yu Dai, Feng-Yu Qiu, Rui Xia, Tao Li, Shu-Jian Huang, and Jia-Jun Chen, 2016. Collaborative Filtering with Topic and Social Latent Factors Incorporating Implicit Feedback.}

\begin{bottomstuff}
Author's addresses: G.-N. Hu, X.-Y. Dai, F.-Y. Qiu, S.-J. Huang, {and} J.-J. Chen are with the National Key Laboratory for Novel Software Technology, Nanjing University, Nanjing 210023, China; emails: njuhgn@gmail.com, \{daixinyu,chenjj\}@nju.edu.cn, \{qiufy,huangsj\}@nlp.nju.edu.cn; R. Xia is with the School of Computer Science and Engineering, Nanjing University of Science and Technology, Nanjing 210023, China.; email: rxia@njust.edu.cn; T. Li is with the School of Computer Science \& Technology, Nanjing University of Posts and Telecommunications (NJUPT), Nanjing, 210046, P.R. China, and the School of Computing and Information Sciences, Florida International University, Miami, FL 33199; email: taoli@cs.fiu.edu. \protect \\
*Corresponding Author: Xin-Yu Dai (daixinyu@nju.edu.cn)
\end{bottomstuff}

\maketitle

\section{Introduction}

For all the benefits of the information abundance and communication technology, the ``information overload'' is one of the digital-age dilemmas we are confronted with. Recommender systems (RSs) are instrumental in tackling this problem~\cite{sarwar01:itemCF,linden03:amazon,next05,PMF,koren09:MF}. They help offer potential interesting information to individual consumers and allow online users to quickly find the personalized information that fits their needs. RSs are nowadays ubiquitous in various domains and e-commerce platforms. They are used to recommend books at Amazon.com~\cite{linden03:amazon}, music at Last.fm and Pandora.com, movies at Netflix.com~\cite{netflix07}, videos at YouTube.com, and resources in scientific community~\cite{CoFact11}.

{\em Collaborative filtering} (CF) approaches play a central role in traditional recommender systems, which are extensively investigated in research community and widely used in industry~\cite{sarwar01:itemCF,linden03:amazon,next05,CFFnT11}. They are based on the simple intuition that if users rated items similarly in the past, then they will be likely to rate other items similarly in the future. Latent factors CF, which learns a latent vector of preferences for each user and a latent vector of attributes for each item, gains popularity and becomes the standard model for recommender due to its accuracy and scalability~\cite{PMF,SVDPP,koren09:MF}. CF models, however, suffer from data sparsity and the imbalance of ratings. They perform poorly on cold users and cold items for which there are no or few data.

To overcome these weaknesses, additional sources of information are integrated into RSs. Social networking and knowledge sharing sites like Twitter.com, Facebook.com, Epinions.com, and Ciao.com are popular platforms for users to connect to each other, to participate in online activities, and to generate shared content or opinions~\cite{li2015recommending}. Social relations~\cite{SoRec,tranMF,DecTrust,LOCABAL,TrustSVD} and item reviews~\cite{jakob09review,ganu09,CTR,HFT,TopicMF} provide independent and diverse sources for recommendation beyond the explicit rating information source, which present both opportunities and challenges for traditional RSs. One research thread, which we call {\em topic matrix factorization} (Topic MF), is to integrate ratings with item contents or reviews text~\cite{jakob09review,ganu09,CTR,HFT,TopicMF}. Reviews justify the rating behavior of a user; ratings are associated with item attributes which can be revealed from reviews. Topic MF methods combine latent factors in ratings with latent topics in item reviews. Another research thread, which we call {\em social matrix factorization} (Social MF), is to combine ratings with social relations~\cite{SoRec,trustEnsemble,tranMF,DecTrust,LOCABAL,TrustSVD}. Extensive studies have found higher likelihood of establishing social ties among people having similar characteristics, namely the theory of homophily, and homophilous ties become effective means of social influence~\cite{homophilyTrust}. Social MF methods factorize rating matrix and social matrix simultaneously assuming that both matrices share the common latent user space. Nevertheless, both Social MF and Topic MF ignore some useful information, either item reviews or social relations.

Contrast to integrating more data sources into RSs, another way is to mine the limited data information more deeply. For example, the SVD++~\cite{SVDPP} model factorizes only the rating matrix and exploits the implicit feedback from it. Item ratings tell us \textit{how} a user rated an item, i.e., an explicit rating score (e.g. from 1 to 5) to indicate her preference degree. Moreover, implicit feedback is associated with these explicit rating scores, telling us \textit{which} items the user rated. Users chose to indicate their preferences implicitly by voting a rating, leaving the rating high or low alone. In another way, users who have rated the same/ similar set of items are more likely to have similar preferences than those who have not, in an a priori sense. The TrustSVD~\cite{TrustSVD} model employs rating and social relation data sources and exploits the implicit feedback from both of them. These methods mines the limited data sources deeply and achieves good results comparing to the shallow mining.

A recent work~\cite{CCTRSoRec} adopts CTR ({\em collaborative topic regression}, one of the Topic MF methods)~\cite{CTR} to exploit ratings and reviews, and adopts SoReg ({\em social regularization}, one of the Social MF methods)~\cite{SoReg} to exploit ratings and social relations. Experimental results show performance improvements compared to the two individual components. Similar methods are also proposed for tag recommendation~\cite{CTRSR} and article recommendation~\cite{CTRSoRec}. These methods, however, have two limitations. First, the two components they used are not effective. For example, some improved components were proposed like HFT ({\em hidden factors and topics}, one of the Topic MF methods)~\cite{HFT} to exploit item reviews and LOCABAL ({\em local and global}, one of the Social MF methods)~\cite{LOCABAL} to exploit social relations. Second, they did not mine the data sources more deeply like incorporating implicit feedback from ratings, though RSs can indeed benefit from implicit information; and this benefit has already been demonstrated like in the SVD++~\cite{SVDPP,koren09:MF} and the TrustSVD~\cite{TrustSVD} models. In this paper, we attempt to overcome these two drawbacks.

\textbf{Contributions.\quad}
First, we investigate the effectiveness of fusing social relations and reviews to rating prediction in a novel way, inspired by the complementarity of the two independent sources for recommendation. The core idea is the alignment between latent factors found by Social MF and hidden topics found by Topic MF to form a unified model. Through latent factors and hidden topics we can learn the representations of user entity and item entity from heterogenous data sources; and the connections among data sources can be reflected in the dependencies among these representations. In this way, we can gain the maximum benefits from all of the information effectively.

Second, we mine the data more deeply by incorporating implicit feedback from ratings into the proposed model to enhance its capability and to demonstrate its flexibility. The core idea is to learn an extra implicit feature matrix to consider the influence of rated items. Due to the sparseness of data, users who have few data will have latent features close to the average, leading to their predicted ratings close to the items' average. Through the implicit features, the users' rated items will have an a priori impact on their ratings with respect to unseen items.

Our main contributions are outlined as follows.
\begin{itemize}
\item {Proposing a novel model {\em MR3} to jointly model user-item ratings, social network structure, and item reviews for rating prediction and along with an extended Social MF method which exploits the ratings and social relations more tightly by capturing the graph structure of neighbors; the MR3 model integrates two effective components.} (Section~\ref{paper:M3R})
\item {Extending the proposed model to obtain a new model {\em MR3++} by incorporating implicit feedback from ratings to enhance its capability and to demonstrate its flexibility; the extension model mines the limited information more deeply by introducing implicit features which captures the influence of rated items.} (Section~\ref{paper:M3R++})
\item {Adapting an alternating optimization algorithm to learn the proposed models which contain different kinds of parameters.} (Section~\ref{paper:learning})
\item {Evaluating the proposed models extensively on two real-world datasets to demonstrate their performance and to understand their working.} (Section~\ref{paper:Exp})
\end{itemize}

A preliminary version of the work, i.e. MR3, has been published in~\cite{MR3}. In this journal submission, we have extended our previous conference paper from the following aspects. First, we propose a new model MR3++ by incorporating implicit feedback from ratings to mine the limited information more deeply (Section~\ref{paper:M3R++}). Second, we refine the preliminaries (Section \ref{paper:preliminary}) and abstract the learning processing (Section \ref{paper:learning}) which are suitable for both MR3 and MR3++. Third, we evaluate the newly proposed model extensively on two real-world datasets, including: (i) prediction performance (Section~\ref{paper:Exp-M3R-RR}), (ii) impact of implicit feedback (Section~\ref{paper:impact-M3R-RR}) and its comparison with contribution of more data sources (Section~\ref{paper:more-vs-deep}), and (iii) the hyperparameters analysis (Section~\ref{paper:mr3pp-hyper-parameters}). Finally, we give further analysis of contributions from auxiliary sources of data (Section~\ref{paper:contribution-further-anlysis}).

The organization of this paper is as follows. We first review related works in Section~\ref{paper:Work}. We introduce the notation and preliminaries in Section \ref{paper:preliminary}. We present details of the proposed model in Section \ref{paper:M3R}, and its extension to incorporate implicit feedback in Section \ref{paper:M3R++}. Their learning processes are described in Section \ref{paper:learning}. In Section \ref{paper:Exp}, we demonstrate our methods empirically on two real-life datasets. We give concluding remarks and discussion of some future work in Section~\ref{paper:conclusion}.

\section{Related works}\label{paper:Work}

We review some related works on collaborative filtering based recommender systems, which are divided into four categories according to information sources they exploit among ratings, reviews, and social relations. Implicit feedback from ratings is also reviewed in the corresponding category.

\textbf{Collaborative Filtering.\quad} {\em Collaborative filtering} (CF) recommender approaches have two types: memory-based CF and model-based CF~\cite{next05,CFFnT11}. They both assume that if users rated items similarly in the past, then they will be likely to rate other items similarly in the future. Memory-based CF methods are grouped into user-based CF and item-based CF. The former predicts the rating of an active user based on the ratings of other similar users on the item~\cite{CFEmpirical}, while the latter based on the ratings of other similar items given by the same user~\cite{sarwar01:itemCF,linden03:amazon}. And the two kinds of memory-based CF can be unified by similarity fusion in a generative probabilistic model~\cite{user-item-CF06}. The similarity between two users or two items can be measured by Pearson correlation or cosine similarity from past rating history. Model-based CF learns a model from past ratings and then uses the learned model to predict unseen ratings. Latent semantics models~\cite{latentSemanticCF04} are some early representative model-based CF. And latent factors CF or matrix factorization based CF, which learns a latent vector of preferences for each user and a latent vector of attributes for each item, gains popularity and becomes the standard model for recommender due to its accuracy and scalability~\cite{koren09:MF,PMF}. These matrix factorization (MF) techniques include maximum margin MF~\cite{fastMMMF05}, nonnegative MF~\cite{wNMF10}, and collective MF~\cite{CollectiveMF}. A relaxation on the assumption that interprets ratings as numerical values is to interpret them as ordinal ones~\cite{ordinalCF}.

Implicit feedback from ratings can readily be added into the basic matrix factorization based CF models~\cite{SVDPP,PMF,FM}. The intuition behind the implicit information from ratings is that users who have rated the same/similar items are more likely to have similar preferences than those who have not, in an a priori sense. Another way to exploit implicit feedback from ratings is to deduce the preference-confidence pairs from the raw observed ratings~\cite{CFImplicit08}. The observed rating data is treated as an indication of positive and negative preferences associated with varying confidence levels. The resulting objective function sums over all the full matrix entries rather than over the only observed ones~\cite{unifyExpImp10}. CF models, however, suffer from data sparsity and the imbalance of ratings; and they perform poorly on cold users and cold items for which there are no or few data. Currently, there are mainly two threads to alleviate these problems: topic matrix factorization integrating reviews text information and social matrix factorization integrating social network information.

\textbf{Topic Matrix Factorization.\quad} One research thread, which we call {\em topic matrix factorization} (Topic MF), is to integrate ratings with item contents or reviews text~\cite{CTR,CTPF,HFT,RMR,TopicMF}. Early works~\cite{jakob09review,ganu09} extract the fine-grained ratings of item aspects from online reviews and then use them as content-based features for collaborative filtering. Aspects of a movie have actors, genres and visual effects; aspects of a restaurant have price, cleanness and service; and aspects of a hotel have location, cleanliness, and room view. The extraction of aspects needs domain knowledge and some amount of manual interaction. Besides item reviews, topic modeling approaches have been incorporated into recommender systems to find the hidden topics of documents like scientific articles~\cite{LDA,CTR,CTPF}. Such methods are belonging to One-Class CF~\cite{OneCF} where the dimensions they discover are not correlated with ratings.

In recent works, some authors adopt {\em latent Dirichlet allocation (LDA)}~\cite{HFT} or {\em nonnegative matrix factorization (NMF)}~\cite{TopicMF} to learn latent topic factors from item reviews and meanwhile adopt a matrix factorization model to exploit the ratings. To bridge the topic-specific factors and rating-specific factors, softmax transformations/exponential functions are proposed to link the two. These methods assume that the dimensionality discovered in ratings is the same as that found in reviews. Ling et al~\cite{RMR} replaced the matrix factorization model with a mixture of Gaussian to avoid the difficult choice of the transformations; Xu et al~\cite{CMR} used the co-clustering of user community and item group to generate the rating distributions and topic distributions, allowing the different dimensionality of user factors and item factors; and Diao et al~\cite{JMARS} introduced the aspect-based model to link the interest distribution of users and the content distribution for movies, allowing the dimensionality of the two to be different. In general, Topic MF methods combine latent factors in ratings with latent topics in item reviews. Nevertheless, Topic MF ignores some useful information, e.g., social relations.

\textbf{Social Matrix Factorization.\quad} Another research thread, which we call {\em social matrix factorization} (Social MF), is to combine ratings with social relations~\cite{SoRec,SoReg,tranMF,LOCABAL,DecTrust,TrustSVD}. Early work~\cite{trustAware07} computed the trust value between users from the trust network to replace the similarity weight in user-based CF. Empirical results show that recommender systems incorporating trust information are effective in terms of accuracy while alleviating the cold-user problem. It's a memory-based CF or trust-related neighborhood model and doesn't study the rating matrix and the trust network systematically.

In recent works, the authors~\cite{SoRec,PrSocRec,LOCABAL,TrustMF,DecTrust,TrustSVD} factorize the rating matrix and the social matrix simultaneously assuming that they share the common latent user space. They exploit the social relations from multiple views: 1) local and global (LOCABAL) view~\cite{LOCABAL} where {\em local} perspective reveals the correlations between the user and her local neighbors while {\em global} perspective indicates the reputation of users in the global network; 2) trustee and truster (TrustMF) view~\cite{TrustMF} where the {\em trustee} model captures how others follow the rating of a user while the {\em truster} model captures how other users affect the rating of a user; and 3) decomposed trust view~\cite{DecTrust} where four trust aspects (benevolence, integrity, competence, and predictability) are formulated to predict the trust values and the predicted trust is combined with the similarity of the latent user feature vectors to get the total trust between two users.

Another way to exploit social network information is to use it as a {\em social regularization (SoReg)} which constrains that the latent factors of users should be close to the average of their trusted neighbors~\cite{SoReg}. A similar work to this is also proposed in~\cite{tranMF} to allow trust propagation. Nevertheless, Social MF ignores some useful information, e.g., reviews text.

\textbf{Hybrid Recommender Approaches.\quad} There is a tendency towards hybrid approaches. The authors in~\cite{CoFact11} propose the matrix co-factorization techniques to exploit the user and the item side information for one-class collaborative filtering~\cite{OneCF}; their objective is to minimize the reconstruction loss of both user-word and item-word TFIDF weight matrix. A recent work to exploit the three types of information for recommendation~\cite{CCTRSoRec} adopts CTR ({\em collaborative topic regression}, one of the Topic MF methods)~\cite{CTR} to exploit ratings and reviews, and adopts SoReg ({\em social regularization}, one of the Social MF methods)~\cite{SoReg} to exploit ratings and social relations. Experimental results show better performance compared to the two individual components. Similar methods are also proposed for tag recommendation~\cite{CTRSR}, celebrity recommendation~\cite{celebrityCTRSocial}, and article recommendation~\cite{CTRSoRec}. However, these models have two drawbacks. First, the two components they used are not effective and some better components were proposed like the HFT ({\em hidden factors and topics}, one of the Topic MF methods) model~\cite{HFT} and the LOCABAL ({\em local and global}, one of the Social MF methods) model~\cite{LOCABAL}. Second, they did not mine the data sources more deeply like exploiting implicit feedback from ratings, though RSs can indeed benefit from implicit information as demonstrated by the SVD++ model and the TrustSVD model. In this paper, we attempt to overcome these two drawbacks.

\begin{figure}
\centering
\includegraphics[height=5.0cm,width=3.2in]{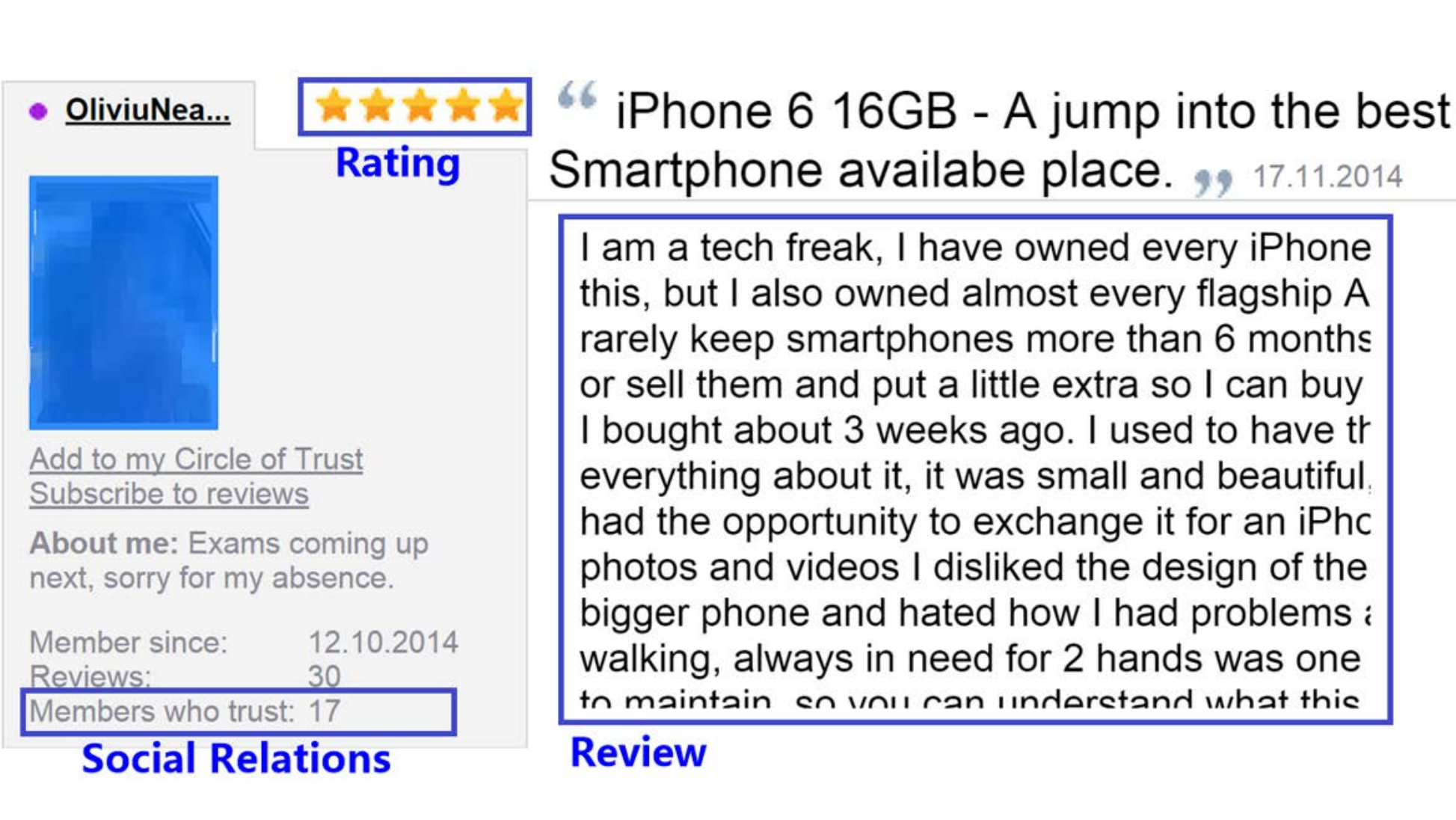}
\caption{ {\em Three types of recommendation information sources.} On the Ciao and Epinions datasets, there are three kinds of data sources, i.e., rating scores, item reviews, and social relations. }
\label{fig:CiaoExample}
\end{figure}

\section{Preliminaries} {\label{paper:preliminary}}

Before proposing our models, we review briefly the representative approaches to exploit the three types of information individually. To this end, we first introduce the notations related to the three data sources shown in Figure~\ref{fig:CiaoExample}.

\textbf{Notations.\quad}
Suppose there are $M$ users $\mathcal{P}=\{u_1,...,u_M\}$ and $N$ items $\mathcal{Q}=\{i_1,...,i_N\}$. We reserve $u$,$v$,$w$ to index the users, and $i$,$j$,$k$ to index the items. Let $R \in \mathbb{R}^{M \times N}$ denote the rating matrix, where the entry $R_{u,i}$ is the rating of user $u$ on item $i$, and we mark a zero if it is unknown. The task of rating prediction is to predict the unknown/missing ratings from the observed data.

In addition to this explicit rating information, other side information sources may exist. One such source is social relations. Users connect to others in a social network, where a link between two users indicates their friendship or trust relation. We use $T \in \mathbb{R}^{M \times M}$ to indicate the user-user social relations, where the entry $T_{u,v}$ = 1 if user $u$ has a relation to user $v$ or zero otherwise. Another side data source is item reviews. Items have affiliated content information, e.g., reviews commented by users. The observed review data $d_{u,i}$ is a piece of text of item $i$ written by user $u$, often along with a rating score $R_{u,i}$.\footnote{We organize these reviews as a document-term matrix $w \in \mathbb{N}^{N \times L}$ where the entry $w_{d,n}$ is the occurrence of token $n$ in doc $d$. For convenience, we also refer the review as $d_{u,i}$.}

Notations used throughout the paper are summarized in Table~\ref{table:notation}.

\textbf{Rating Information.\quad}
For the information source of ratings, matrix factorization based latent factor models~\cite{PMF,koren09:MF} are mainly to find the latent user-specific feature matrix $P=[P_1,...,P_M] \in \mathbb{R}^{F \times M}$ and item-specific feature matrix $Q=[Q_1,...,Q_N] \in \mathbb{R}^{F \times N}$ to approximate the observed rating matrix in the least-squares sense (more precisely, regularized least squares or ridge regression), obtained by solving the following problem
\begin{equation}
\label{eq:rating}
\min_{P,Q} \sum\nolimits_{R_{u,i} \neq 0} {(R_{u,i} - \hat R_{u,i})}^2 + \lambda (\norm {P}_{Fro}^2 + \norm {Q}_{Fro}^2),
\end{equation}
where $\lambda$ is the regularization parameter to avoid over-fitting and the predicted ratings, and $\norm{.}_{Fro}^2$ denotes the Frobenius norm.
\begin{equation}
\label{eq:pred}
\hat R_{u,i} = \mu + b_u + b_i + P_u^{\textrm T} Q_i.
\end{equation}
The parameters $\mu$, $b_u$ and $b_i$ are the mean of ratings, bias of the user and bias of the item, respectively. The $F$-dimensional feature vectors $P_u$ and $Q_i$ represent users' preferences and items' characteristics. The dot products capture their interaction or match degree.

\textbf{Social Information.\quad}
For the information source of social relations, social matrix factorization methods~\cite{ContentLink07,LOCABAL} are mainly to find the latent social-specific feature matrix $P$ and social correlation matrix $H \in \mathbb{R}^{F \times F}$ to approximate the observed social similarity matrix $S \in \mathbb{R}^{M \times M}$ in the least-squares sense, by solving the following problem \footnote{We omit the term about the ratings, which is the same as that in Eq.(\ref{eq:rating}), to show clearly how to exploit the social relations.}
\begin{dmath}
\label{eq:locabal}
\min_{P,H} \sum\nolimits_{T_{u,v} \neq 0} {(S_{u,v} - P_u^{\mathrm{T}} H P_v)}^2 + \lambda (\norm {P}_{Fro}^2 + \norm{H}_{Fro}^2),
\end{dmath}
where $S_{u,v}$ is the social similarity between user $u$ and her trustee $v$, defined as the cosine similarity between their rating vectors
\begin{dmath}
\label{eq:socialSimilarity}
S_{u,v} = \sum\nolimits_i {R_{u,i} \cdot R_{v,i}} \Big{/} \sqrt{\sum\nolimits_i {R_{u,i}^2} \cdot \sum\nolimits_i {R_{v,i}^2}}.
\end{dmath}
The assumption to make the above method work is that the latent social-specific matrix is shared with the latent user-specific matrix; both are referred as $P$ here. Namely, users have dual identity: one is involved in rating behavior and the other is involved in social behavior.

\begin{table}
\centering
\tbl{{Notations}\label{table:notation}}{
\begin{tabular}{l l l}
 \Xhline{2\arrayrulewidth}
 Symbol & Meaning & Form \\
 \hline  \hline
 $M$, $N$   & the number of users, and of items & scalar \\
 $L$        & the size of the word vocabulary & scalar \\
 $\mathcal{P}$, $\mathcal{Q}$  & the set of users, and of items & set \\
 $N_d$      & the set of words in doc $d$ & set \\
 $N_u$      & the set of items rated by the user $u$ & set\\
 $T_u$      & the set of users trusted by the user $u$ & set\\
 \hline
 $R_{u,i}, R^b_{u,i}$  & rating of item $i$ by user $u$, and its implicit binary rating &  $R \in \mathbb{R}^{M \times N}$, $R^b \in {\{0,1\}}^{M \times N}$\\
 $T_{u,v}$  & social relation between user $u$ and $v$ & $T \in \mathbb{N}^{M \times M}$\\
 $w_{d,n}$  & the $n^{\mathrm{th}}$ word in doc $d$ & doc-term matrix, $w \in \mathbb{N}^{N \times L}$ \\
 \hline
 $W_{u,i}$  & weight on the rating of item $i$ given by user $u$ & pre-computed, $W \in \mathbb{R}^{N \times N}$\\
 $C_{u,v}$  & social strength (trust value) between user $u$ and $v$ & pre-computed, $C \in \mathbb{R}^{M \times M}$\\
 $S_{u,v}$  & social similarity between user $u$ and $v$ & pre-computed, $S \in \mathbb{R}^{M \times M}$\\
 \hline
 $F$        & dimensionality of latent factors/topics & hyper-parameter, scalar \\
 \hline
 $P_u$      & $F$-dimensional feature vector for user $u$ & parameters, $P \in \mathbb{R}^{F \times M}$\\
 $Q_i$      & $F$-dimensional feature vector for item $i$ & parameters, $Q \in \mathbb{R}^{F \times N}$\\
 $Y_j$      & $F$-dimensional implicit feature vector for item $j$ & parameters, $Y \in \mathbb{R}^{F \times N}$\\
 $\theta_i$ & $F$-dimensional topic distribution for item $i$ & parameters, $\theta \in \Delta^{F \times N}$\\
 $\phi_f$, $\psi_f$   & word distribution for topic $f$, and the unnormalized one & parameters, $\phi \in \Delta^{L \times F}$\\
 $H$        & social correlation matrix & parameters, $H \in \mathbb{R}^{F \times F}$ \\
 \Xhline{2\arrayrulewidth}
\end{tabular}}
\end{table}

\textbf{Review Information.\quad}
For the information source of reviews, topic modeling approaches are used to find the item properties and hidden topics~\cite{LDA,HFT}. The negative log-likelihood (NLL) of the reviews collection is defined as~\footnote{We aggregate all reviews of a particular item as a `doc'; so the indices of docs are corresponding to those of  items.}
\begin{equation}
\label{eq:reviewLikelihood}
- \sum\nolimits_{d=1}^N \sum\nolimits_{n \in N_d} \Big(\log \theta_{z_{d,n}} + \log \phi_{z_{d,n},w_{d,n}} \Big),
\end{equation}
where the parameters $\theta$ and $\phi$ are the topic and word distributions, respectively; $w_{d,n}$ and $z_{d,n}$ are the word and the corresponding topic in doc $d$. Reviews explain the ratings why the users rate in that way. Intuitively, a certain property of the item is probably discussed by a specific distribution of words which corresponds to a certain topic. The following softmax transformation sharpens this intuition and also bridges the gap between the real-valued parameters $Q_i \in \mathbb{R}^F$ associated with ratings and the corresponding probabilistic ones $\theta_i \in \Delta^F$ associated with reviews.
\begin{dmath}
\label{eq:tran}
\theta_{i,f} = \exp (\kappa Q_{i,f}) \Big{/} \sum\nolimits_{f=1}^F \exp (\kappa Q_{i,f}),
\end{dmath}
where the parameter $\kappa$ controls `peakiness' of the transformation and can be jointly learned with other parameters.

\begin{figure}
\centering
\includegraphics[height=3.8cm,width=4.4in]{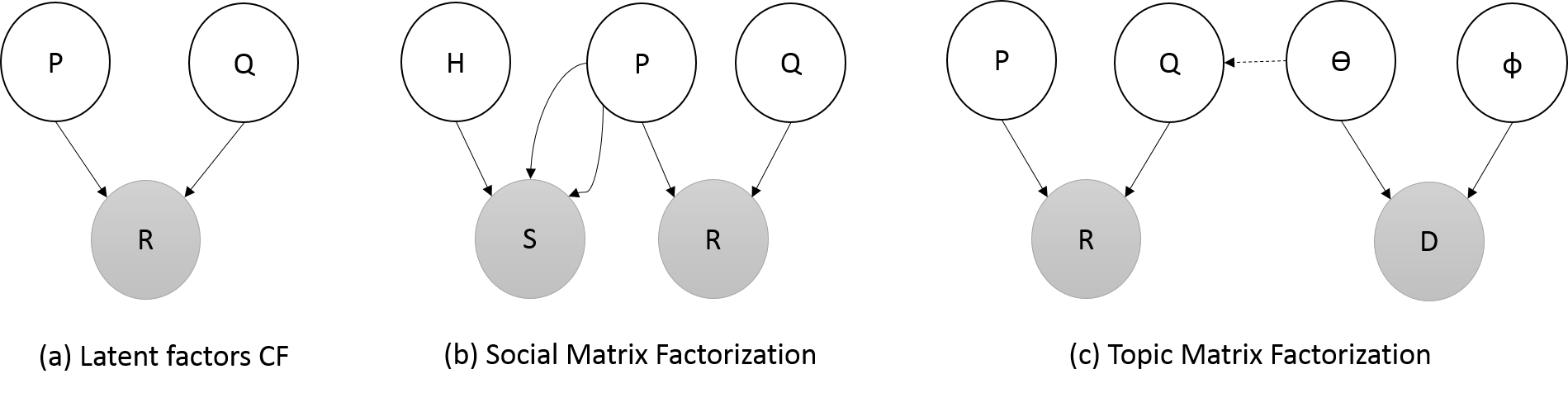}
\caption{ {\em Illustrations of the dependencies among data matrices and parameter matrices  in three different recommendation approaches.} (a) Latent factors CF exploits ratings, (b) Social MF integrates ratings with social relations; and (c) Topic MF integrates ratings with reviews text. Shaded nodes are data and others are parameters.}
\label{fig:preliminary}
\end{figure}

The dependencies among data matrices and parameter matrices in three different recommendation approaches are shown in Figure~\ref{fig:preliminary}. The left subfigure is latent factors CF approach to exploit the rating information; the middle subfigure is social matrix factorization approach to integrate ratings with social relations; and the right subfigure is topic matrix factorization approach to combine ratings with item reviews.

\section{The Proposed Model}\label{paper:M3R}

In this section, we propose a model to solve the following problem. We call it Problem 1 which requires to model three data sources simultaneously.

\subsection{Problem 1}\label{paper:problem1}

\theoremstyle{problem}
\begin{problem}
Rating Prediction with Social Relations and Reviews.

\textbf{Input:} 1) a rating matrix $R$, 2) a social network among users $T$, 3) a reviews collection along with the ratings $D$, 4) a user $u$ in the user set $\mathcal{P}$, and 5) an item $i$ in the item set $\mathcal{Q}$.

\textbf{Output:} the predicted preference of user u on item i, where $u \in \mathcal{P}$ and $i \in \mathcal{Q}$.
\end{problem}

In Problem 1, to predict the unknown ratings, we have three types of information to exploit: ratings, social relations, and item reviews.

\subsection{MR3: A Model of Ratings, Item Reviews and Social Relations}{\label{paper:m3r}}

In Section~\ref{paper:preliminary}, we have reviewed the three kinds of approaches to exploit the three kinds of information sources individually (see Figure~\ref{fig:preliminary}), i.e., matrix factorization based collaborative filtering for the information source of ratings (see Eq.(\ref{eq:rating}), or Figure~\ref{fig:preliminary}(a)), social matrix factorization for the information source of social relations (see Eq.(\ref{eq:locabal}), or Figure~\ref{fig:preliminary}(b)), and the topic matrix factorization for the information source of item reviews (see Eq.(\ref{eq:reviewLikelihood}), or Figure~\ref{fig:preliminary}(c)). With these preliminaries in mind, we can present our solution to the Problem 1, exploiting all of data sources simultaneously.

For Problem 1, the main challenge is how to fuse effectively the three heterogenous data sources to form a unified model. We tackle this challenge by combining the two parts, i.e. Social MF and Topic MF, described in the following. The core idea is, based on the collaborative filtering, the alignment between latent factors and hidden topics found by the above two parts.

In one part, the LOCABAL ({\em local and global}, one of the Social MF methods) model~\cite{LOCABAL} exploits the ratings and social relations by incorporating social latent factors into collaborative filtering. The goals to achieve are modeling ratings accurately and also capturing the social context, by solving the following problem:~\footnote{The global social context $W_{u,i}$ is omitted here for clarity, and given in Eq.(\ref{eq:m3r}) instead.}
\begin{dmath}
\label{eq:socialMF}
\min_{P,Q,H} \sum\nolimits_{R_{u,i} \neq 0} W_{u,i}{(R_{u,i} - \hat R_{u,i})}^2 + \lambda_{\mathrm{rel}} \sum\nolimits_{T_{u,v} \neq 0} {(S_{u,v} - P_u^{\mathrm{T}} H P_v)}^2 + \lambda \Omega (\Theta),
\end{dmath}
where the rating weights $ \label{eq:W} W_{u,i} = 1/(1+\log r_u) $ are computed from the PageRank scores of users in the social network, representing the global perspective of social context~\cite{LOCABAL}, where $r_u$ is the rank of user $u$ in decreasing order, i.e, the top-ranked users having high ranking scores. The regularization parameter $\lambda_{\mathrm{rel}}$ controls the contribution from social relations, parameters $\Theta = \{P,Q,H\}$, and the regularization term is given by $\Omega (\Theta)=\norm {P}_{Fro}^2 + \norm {Q}_{Fro}^2 + \norm{H}_{Fro}^2$.

In another part, the HFT ({\em hidden factors and topics}, one of the Topic MF methods) model~\cite{HFT} exploits the ratings and item reviews by incorporating topic latent factors into collaborative filtering. The goals to achieve are both modeling ratings accurately and generating reviews likely, by solving the following problem:
\begin{dmath}
\label{eq:topicMF}
\min_{P,Q,\Phi} \sum\nolimits_{R_{u,i} \neq 0} {(R_{u,i} - \hat R_{u,i})}^2 - \lambda_{\mathrm{rev}} \sum\nolimits_{d=1}^N \sum\nolimits_{n \in N_d} \Big(\log \theta_{z_{d,n}} + \log \phi_{z_{d,n},w_{d,n}} \Big),
\end{dmath}
where $\lambda_{\mathrm{rev} }$ controls the contribution from reviews, and parameters $\Phi = \{\theta,\phi\}$. The connection between ratings and reviews is the coupling between parameters $Q_i$ and $\theta_i$ as shown in Eq.(\ref{eq:tran}).

Based on collaborative filtering, we can exploit the three data sources simultaneously by combining the above two parts (i.e., Eq.(\ref{eq:socialMF}) and Eq.(\ref{eq:topicMF})). By aligning latent factors and topics, we connect Social MF and Topic MF through the proposed model {\em MR3 (model of rating, relation, and review)}, which minimizes the following problem~\cite{MR3}:
\begin{dmath}
\label{eq:m3r}
\mathcal{L}(\Theta,\Phi,z,\kappa) \triangleq \sum\nolimits_{R_{u,i} \neq 0} W_{u,i} {(R_{u,i} - \hat R_{u,i})}^2 \\
- \lambda_{\mathrm{rev}} \sum\nolimits_{d=1}^N \sum\nolimits_{n \in N_d} \Big(\log \theta_{z_{d,n}} + \log \phi_{z_{d,n},w_{d,n}} \Big) \\
+ \lambda_{\mathrm{rel}} \sum\nolimits_{T_{u,v} \neq 0} C_{u,v} {(S_{u,v} - P_u^{\textrm T} H P_v)}^2 + \lambda \Omega (\Theta) ,
\end{dmath}

\begin{figure}
\centering
\includegraphics[height=5.0cm,width=3.6in]{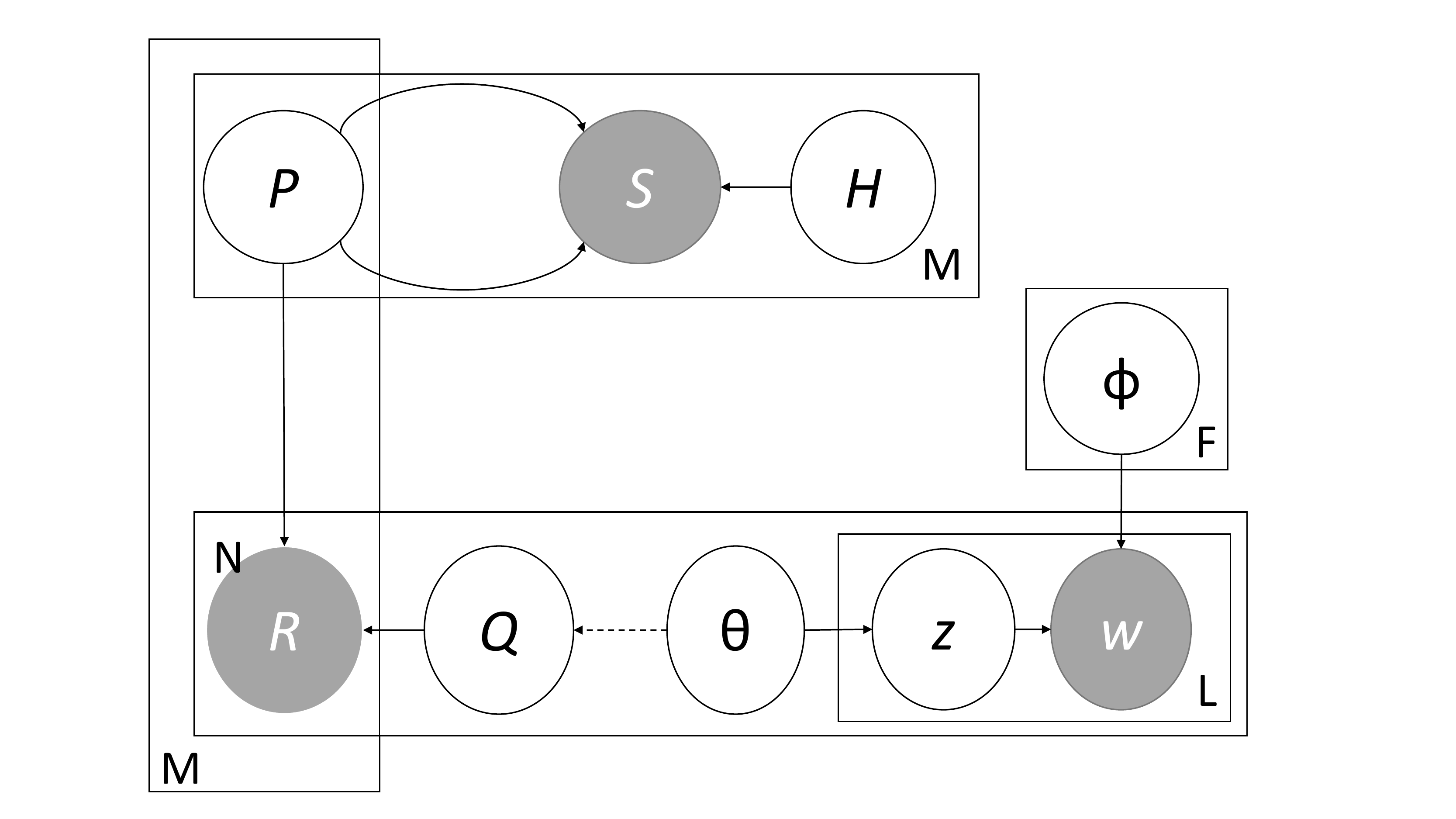}
\caption{ {\em Relationship among matrices of parameters and data of the proposed model (MR3).} Shaded nodes are data ($R$: rating matrix, $S$: social similarity matrix, and $D$ (here it is embodied by words $w$): doc-term matrix of reviews). Others are parameters ($P$: matrix of latent user factors, $Q$: matrix of latent item factors, $H$: social correlation matrix, $\theta$: doc-topic distributions, and $\phi$: topic-word distributions). The double connections between $P$ and $S$ are indicated by the term $(S - P^{\textrm T}HP)$ in Eq.(\ref{eq:locabal}). The dotted line between $Q$ and $\theta$ indicates their coupling through the fixed transformation as shown in Eq.(\ref{eq:tran}). }
\label{fig:depend}
\end{figure}

Beyond the LOCABAL model, we borrow the idea {\em trust values} from the SoRec method~\cite{SoRec} to exploit the ratings and social relations more tightly
\begin{equation}
\label{eq:trustValues}
C_{u,v} = \sqrt{d^-_{v} / (d^+_{u} + d^-_{v})},
\end{equation}
where the outdegree $d^+_{u}$ represents the number of users whom $u$ trusts, while the indegree $d^-_{v}$ denotes the number of users who trust $v$. The trust values are used to capture the graph structure of neighbors, representing the social influence locality, i.e., user behaviors are mainly influenced by close/direct friends in their ego networks~\cite{localInfluence},

Before we delve into the learning algorithm, a brief discussion on Eq.(\ref{eq:m3r}) is in order. On the right hand, the first term is rating squared-error weighted by user reputation in the social network; the second term is the negative log likelihood of item reviews corpus; the third term is local social context factorization weighted by trust values among users; and the last term is Frobenius norm penalty of parameters to control over-fitting. The contributions from item reviews and social relations are controlled by the two hyper-parameters $\lambda_{\mathrm{rev}}$ and $\lambda_{\mathrm{rel}}$. The connection between ratings and social relations is through the shared user latent feature space $P$, ratings and reviews are linked through the transformation involving $Q$ and $\theta$ as shown in Eq.(\ref{eq:tran}) where parameter $\kappa$ controls the peakiness of the transformation, and all of the three data sources are jointly modeling based on collaborative filtering with topic and social latent factors. The dependencies among the parameter and data matrices in the proposed model are depicted in Figure \ref{fig:depend}. Note that the dotted line between $Q$ and $\theta$ indicates the fixed transformation between them (see Eq.(\ref{eq:tran})) and does not exhibit any distribution dependency; hence the figure does not describe a truly Bayesian generative model on the whole and the purpose is to clearly display the relationship among matrices of parameters and data.

\noindent
\textbf{eSMF.\quad} We separate the following part from the proposed model MR3 and denote it as the {\em extended Social MF} (eSMF) model~\cite{MR3}
\begin{dmath}
\label{eq:eSMF}
\mathcal{L}(\Theta) \triangleq \sum\nolimits_{R_{u,i} \neq 0} W_{u,i} {(R_{u,i} - \hat R_{u,i})}^2 \\
+ \lambda_{\mathrm{rel}} \sum\nolimits_{T_{u,v} \neq 0} C_{u,v} {(S_{u,v} - P_u^{\textrm T} H P_v)}^2 + \lambda \Omega (\Theta).
\end{dmath}
The eSMF model is the Social MF component of MR3, which extends the LOCABAL model~\cite{LOCABAL} by capturing the graph structure of neighbors via trust values~\cite{SoRec} representing the social influence locality~\cite{localInfluence}.

\section{An Extension of the Proposed Model}\label{paper:M3R++}

In this section, we incorporate the implicit feedback from ratings into the proposed model to enhance its capability and demonstrate its flexibility, leading to a solution to the following problem. We call it Problem 2 which requires to mine the limited information more deeply.

\subsection{Problem 2}\label{paper:problem2}

For item ratings information, they tell us \textit{how} a user rated an item, i.e., an explicit rating score (e.g. from 1 to 5) to indicate her preference degree. Moreover, implicit feedback is always associated with these explicit rating scores, telling us \textit{which} items the user rated. In detail, a binary matrix $R^b$ can be constructed from the rating matrix, where the entry has one if the corresponding one in the rating matrix is observed and zero if it is unseen. Users chose to indicate their preferences implicitly by voting a rating, leaving the rating high or low alone. In another way, users who have rated the same/similar items are more likely to have similar preferences than those who have not, in an a priori sense.

With the above additional consideration, we extend the Problem 1 to reach the following problem.

\theoremstyle{problem}
\begin{problem}
Problem 1 with Implicit Feedback from Ratings.

\textbf{Input:} 1) a rating matrix $R$ and its implicit matrix $R^b$, 2) a social network among users $T$, 3) a reviews collection along with the ratings $D$, 4) a user $u$ in the user set $\mathcal{P}$, and 5) an item $i$ in the item set $\mathcal{Q}$.

\textbf{Output:} the predicted preference of user u on item i, where $u \in \mathcal{P}$ and $i \in \mathcal{Q}$.
\end{problem}

By Problem 2, we define the meaning of ``mine the limited information more deeply'' as ``exploit the implicit feedback constructed from explicit ratings''. That is, besides learning the latent user factors and item factors from ratings, we also learn an implicit feature vector for each item. As we will see in Eq.(\ref{eq:svdpp}), the implicit feedback from ratings can be used as the prior preferences of users.

\subsection{MR3++: An Extension of the Proposed Model Incorporating Implicit Feedback from Ratings}{\label{paper:m3r-implicit}}

In Section~\ref{paper:m3r}, we have introduced the solution to the Problem 1. The solution is the model MR3 (see Eq.(\ref{eq:m3r})), which exploits the three types of information sources simultaneously by connecting the Social MF approach and the Topic MF approach. In this subsection, we extend the proposed model. leading a solution to Problem 2.

\begin{figure}
\centering
\includegraphics[height=5.0cm,width=3.6in]{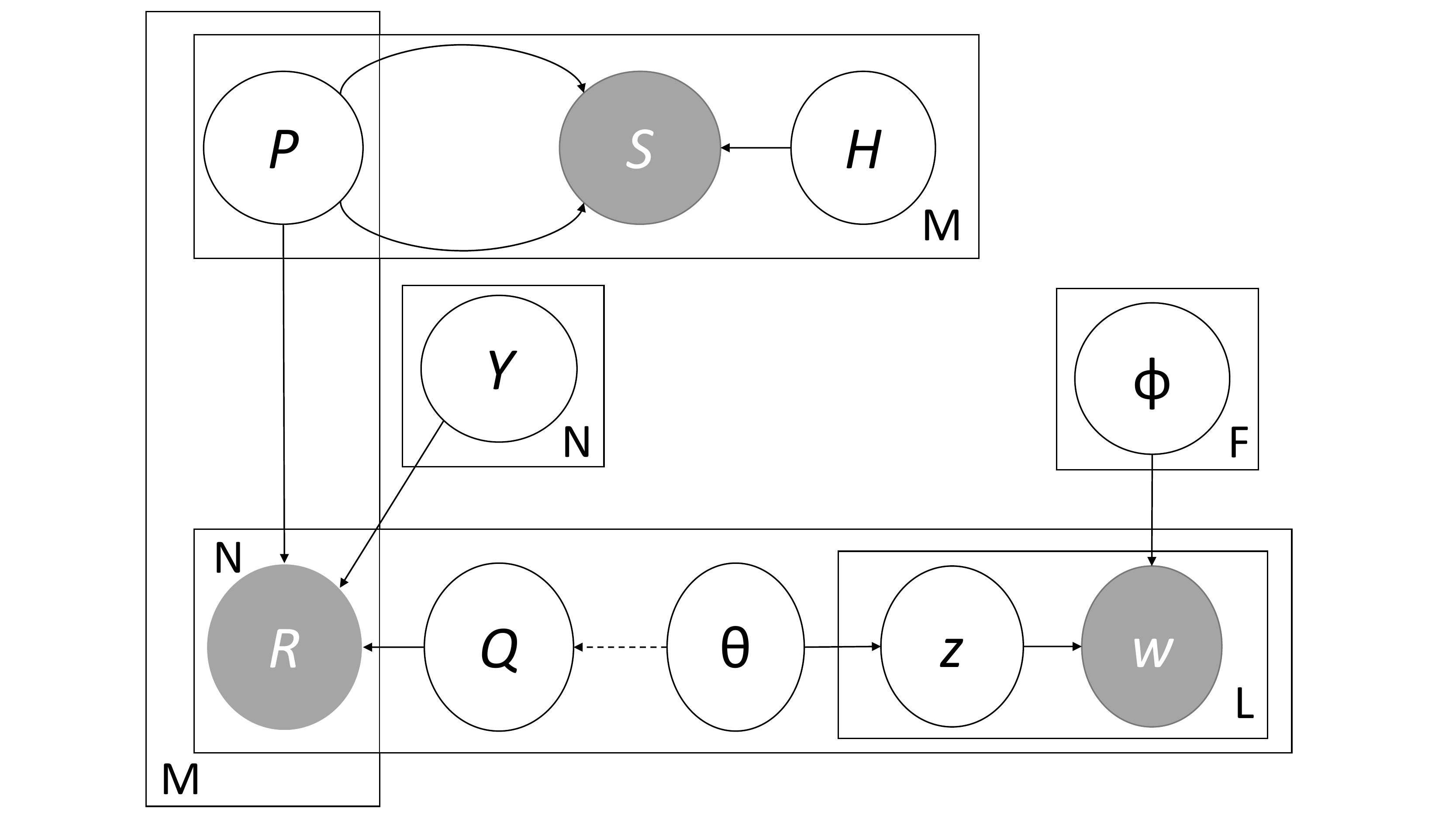}
\caption{ {\em The Extension of the Proposed Model (MR3++).} Shaded nodes are data and others are parameters. Implicit feature matrix $Y \in \mathbb{R}^{F \times N}$ is added to consider the influence of rated items. The total preferences of a user on an item now decompose into two parts: one part indicates `intrinsic preference' reflecting in her latent feature vector; and another part shows the `influence of her rated items' captured by their implicit feature vectors. }
\label{fig:depend-mr3pp-large}
\end{figure}

The core idea of mining ratings deeply is to learn an extra implicit feature matrix $Y \in \mathbb{R}^{F \times N}$ to consider the influence of rated items. Due to the sparseness of data, users who have few data will have latent features close to the average, leading to their predicted ratings close to the items' average. Through the implicit features, the users' rated items will have an a priori impact on their ratings with respect to unseen items. In more detail, the total preferences of user $u$ on item $i$ decompose into two parts: one part indicates some `intrinsic preference' reflecting in her latent feature vector $P_u$; and another part shows the `influence of her rated items' captured by their implicit feature vectors $Y_j$, where the index $j$ denotes the set of the user's rated items. These ideas are shaped in the SVD++ model~\cite{SVDPP} and the constrained PMF model~\cite{PMF}, where the predicted ratings are now computed by (rather than by Eq.(\ref{eq:pred}))
\begin{dmath}
\label{eq:svdpp}
\hat R_{u,i}^* = P_u^{\textrm T} Q_i + \Big( |N_u|^{- \frac 1 \alpha} \sum\nolimits_{j \in N_u} Y_j \Big) ^{\textrm T} Q_i + \mu + b_u + b_i,
\end{dmath}
where $N_u$ is the set of items rated by the user $u$, i.e., $N_u = \{j: R^b_{u,j} = 1$\} and $Y_j$ is the implicit feature vector for item $j$. In the SVD++ model, $\alpha = 2$; in the constrained PMF model, $\alpha = 1$. We can see, for example, if user $u$ has rated the same items as those rated by user $v$, i.e., $N_u = N_v$, then in an a priori sense, these two users are likely to have similar preferences.

The extended model contains two components: one component is to combine three kinds of data sources (ratings, social relations, and reviews), i.e., the MR3 model; and another is to incorporate implicit feedback from ratings and hence mining the rating source more deeply. Hence we call the extended model MR3++, where `++' stands for extending the MR3 model by exploiting implicit feedback from ratings. MR3++ minimizes the following problem:
\begin{dmath}
\label{eq:m3r-ra}
\mathcal{L}(\Theta^*,\Phi,z,\kappa) \triangleq \sum\nolimits_{R_{u,i} \neq 0} W_{u,i} {(R_{u,i} - \hat R_{u,i}^*)}^2 \\
- \lambda_{\mathrm{rev}} \sum\nolimits_{d=1}^N \sum\nolimits_{n \in N_d} \Big(\log \theta_{z_{d,n}} + \log \phi_{z_{d,n},w_{d,n}} \Big) \\
+ \lambda_{\mathrm{rel}} \sum\nolimits_{T_{u,v} \neq 0} C_{u,v} {(S_{u,v} - P_u^{\textrm T} H P_v)}^2 + \lambda \Omega (\Theta^*) ,
\end{dmath}
where $\hat R_{u,i}^*$ is given by Eq.(\ref{eq:svdpp}) (rather than Eq.(\ref{eq:pred})), and
$\Omega (\Theta^*) = \norm {P}_{Fro}^2 + \norm {Q}_{Fro}^2 + \norm {H}_{Fro}^2 + \norm {Y}_{Fro}^2$.

The dependencies among the parameter and data matrices in the extension model are depicted in Figure~\ref{fig:depend-mr3pp-large}, where the implicit feature matrix is added compared to the proposed model.

\textbf{Implicit Feature Matrix $Y$.\quad} Note that the idea of learning feature matrix $Y$ originally (to the best of our knowledge) comes from the NSVD model ~\cite{NSVD} to decrease the number of parameters, where $Y$ is used to replace the learning of latent user feature matrix $P$; hence the complexity of parameters is from $\mathcal{O}(MK + NK)$ to $\mathcal{O}(NK)$ and does not depend on the scale of users. Besides capturing implicit feedback, $Y$ has the effect of learning the item-item similarity $sim(i, j) = Y_j^{\textrm T} Q_i$. The predicted rating is computed by $R_{u,i} = \sum\nolimits_{j \in N_u} sim(i, j) + b_u + b_i$. This method is modified to exclude the known rating information for a given user-item pair when estimating the similarity matrix in the FISM model~\cite{FISM}. The estimated rating is computed as $R_{u,i} = \sum_{j \in N_u\setminus\{i\}} sim(i, j) + b_u + b_i$.

To mine the implicit feedback from ratings, we adopt the SVD++ model idea in our proposed model MR3++: learning both the item-item similarity $sim(i, j) = Y_j^{\textrm T} Q_i$ and the user-item similarity $sim(u, i) = P_u^{\textrm T} Q_i$. And hence the predicted ratings are computed by $R_{u,i} = sim(u, i) + |N_u|^{- \frac 1 2} \sum_{j \in N_u \setminus \{i\}} sim(i, j) + \mu + b_u + b_i$, which is just the same with Eq.(\ref{eq:svdpp}). Combined with the FISM model idea, we can also adopt the following rating predictor in our MR3++ model: \begin{dmath*} R_{u,i} = sim(u, i) + (|N_u| - 1)^{- \alpha} \sum\nolimits_{j \in N_u\setminus\{i\}} sim(i, j) + \mu + b_u + b_i \end{dmath*}, where $\alpha$ is an adjustable hyper-parameter between 0 and 1.

No matter what kind of implicit predictor is used (NSVD, SVD++, or FISM), the idea of ming the limited data more deeply is worth keeping in mind. And this is a meaningful point for extending the MR3 model to the MR3++ model.

\section{Model Learning}{\label{paper:learning}}
We give the optimization algorithms to learn the models (namely MR3 and MR3++) proposed in the above two sections (MR3 in Section~\ref{paper:M3R} and MR3++ in Section~\ref{paper:M3R++}). Their learning processes are the same except the minor difference with respect to gradients of two parameters.

\subsection{Learning Process}{\label{paper:m3r-learning}}

Our objective is to search
\begin{dmath}
\label{eq:m3rLearning}
\argmin_{\Theta,\Phi,z,\kappa} \mathcal{L}(\Theta,\Phi, z,\kappa).
\end{dmath}
Observe that parameters $\Theta=\{P,Q,H\}$ and $\Phi=\{\theta,\phi\}$ are coupled through the transformation between $Q$ and $\theta$ as shown in Eq.(\ref{eq:tran}) (the dotted line shown in Figure \ref{fig:depend}). The former parameters $\Theta$ are associated with ratings and social relations, which can be found by gradient descent methods; while the latter parameters $\Phi$ are associated with reviews text, which can be found by Gibbs sampling~\cite{Gibbs}. So similar to the HFT (hidden factors and topics, one of the Topic MF methods) model [38], we design a procedure alternating between the following two steps:
\begin{subequations}
\label{eq:2step}
\begin{equation}
\label{eq:step1}
 \mbox{update } \Theta^{\mathrm{new}}, \Phi^{\mathrm{new}},\kappa^{\mathrm{new}} = \argmin_{\Theta,\Phi,\kappa} \mathcal{L}(\Theta,\Phi,z^{\mathrm{old}}, \kappa);
 \end{equation}
\begin{equation}
\label{eq:step2}
 \mbox{sample }\; z_{d,n}^{\mathrm{new}} \mbox{ with probability } p(z_{d,n}^{\mathrm{new}} = f) = \phi_{f,w_{d,n}}^{\mathrm{new}}.
\end{equation}
\end{subequations}

For the first step as shown in Eq.(\ref{eq:step1}), topic assignments $z_{d,n}$ for each word in reviews corpus are fixed; then we update the terms {$\Theta,\Phi$, and $\kappa$} by gradient descent (GD). Recall that $\theta$ and $Q$ depend on each other; we fit only $Q$ and then determine $\theta$ by Eq.(\ref{eq:tran}). This is the same as in the standard gradient-based MF for recommender~\cite{PMF} except that we have to compute more gradients, which will be given later separately.

For the second step as shown in Eq.(\ref{eq:step2}), parameters associated with reviews corpus $\theta$ and $\phi$ are fixed; then we sample topic assignments $z_{d,n}$ by iterating through all docs $d$ and each word within, setting $z_{d,n} = f$ with probability proportion to $\theta_{d,f} \phi_{f,w_{d,n}}$. This is similar to updating $z$ via LDA~\cite{LDA} except that topic proportions $\theta$ are not sampled from a Dirichlet prior, but instead are determined in the first step through $Q$.

Finally, the two steps are repeated until a local optimum is reached. In practice, we sample topic assignments every 5 GD epoches and this is called a pass; usually it is enough to run 50 passes to find a local minima. The experimental settings are detailed in the experimental section.

For MR3++, the learning process is the same with that of MR3, except that MR3++ has more gradients to compute and other gradients related to the implicit feature matrix $Y$ have to modify accordingly.
The learning algorithm is summarize in Algorithm~\ref{alg:mr3}.

\begin{algorithm}[b]
\KwIn{Ratings $R$, reviews $w$, and relations $T$, number of latent factors $F$, maximum number of iterations $maxIter$, number of epoches $nEpoch$.}
\KwOut{User features $P$, item features $Q$, social relation matrix $H$, implicit features $Y$, topic proportions $\theta$, and topic distributions $\phi$. }
Pre-compute $W$, $S$, and $C$ by Eq.(\ref{eq:W}), Eq.(\ref{eq:socialSimilarity}), and Eq.(\ref{eq:trustValues})\;
Initialize $P, Q, H, Y$ randomly from $\mathcal{N}(0,0.01)$\;
$iter$ = 1\;
\Repeat{$convergence \quad (e.g., iter > maxIter$ or $ \frac{|\mathcal{L}^{iter} - \mathcal{L}^{iter+1}|}{|\mathcal{L}^{iter}|} < 10^{-6} )$}{
    \For{$epoch = 1; epoch < nEpoch; epoch$ ++}{
         update $\Theta^{\mathrm{new}}, \Phi^{\mathrm{new}},\kappa^{\mathrm{new}} = \argmin_{\Theta,\Phi,\kappa} \mathcal{L}(\Theta,\Phi,z^{\mathrm{old}}, \kappa)$ by SGD using gradients Eqs.(\ref{eq:grad-P} - \ref{eq:grad-Q*})\;
     }

     sample $z_{d,n}^{\mathrm{new}} \mbox{ with probability } p(z_{d,n}^{\mathrm{new}} = f) = \phi_{f,w_{d,n}}^{\mathrm{new}}$ \;

     $iter$ ++\;
}
\caption{Learning process of the proposed models}
\label{alg:mr3}
\end{algorithm}

\subsection{Gradients of the Parameters}{\label{paper:gradients}}
We now give gradients used in Eq.(\ref{eq:step1}). (Gradients of biases are omitted; rating mean is not fitted because ratings are centered before training.) Besides the notations given in Table~\ref{table:notation}, more notations are required~\cite{Gibbs} to learn the models.
\begin{itemize}
\item {For each item $i$ (i.e. doc $i$, generated from all reviews commented on this item): 1) $M_i$ is an $F$-dimensional count vector, in which each component is the number of times each topic occurs for it; 2) $m_i$ is the number of words in it; and 3) $z_i = \sum\nolimits_f \exp{(\kappa Q_{if})}$ is a normalizer.}
\item {For each word $w$ (in the prescribed word vocabulary): 1) $N_w$ is an $F$-dimensional count vector, in which each component is the number of times it has been assigned to each topic; 2) $n_f$ is the number of times topic $f$ occurs for it; and 3)$z_f = \sum\nolimits_w \exp{(\psi_{fw})}$ is a normalizer.}
\end{itemize}

The user-specific feature matrix $P$ exists in three terms: the first is in the rating prediction, the second is in the local social context, and the third is in the Frobenius norm penalty.
\begin{dmath}
\label{eq:grad-P}
 \frac 1 2 \frac {\partial\mathcal{L}} {\partial P_u} = \sum\nolimits_{i:R_{u,i} \neq 0} W_{u,i}(\hat R_{u,i} - R_{u,i})Q_i + \lambda P_u \\
 + \lambda_{\mathrm{rel}} \sum\nolimits_{v:T_{u,v} \neq 0} C_{u,v}(P_u^{\mathrm T}H P_v - S_{u,v}) H P_v \\
 + \lambda_{\mathrm{rel}} \sum\nolimits_{v:T_{v,u} \neq 0} C_{v,u}(P_v^{\mathrm T} H P_u - S_{u,v})H^{\mathrm T} P_v.
\end{dmath}

The item-specific feature matrix $Q$ exists in three terms: the first is in the rating prediction, the second is in the item reviews, and the third is in the Frobenius norm penalty.
\begin{dmath}
\label{eq:grad-Q}
\frac {\partial\mathcal{L}} {\partial Q_i} = 2 \sum\nolimits_{u:R_{u,i} \neq 0} W_{u,i}(\hat R_{u,i} - R_{u,i})P_u \\
 - \lambda_{\mathrm{rev}} \kappa \Big(M_i - \frac {m_i} {z_i} \exp{(\kappa Q_i})\Big) + 2 \lambda Q_i.
\end{dmath}

The social correlation matrix $H$ exists in two terms: one is in the local social context, and the other is in the Frobenius norm penalty.
\begin{dmath}
\label{eq:grad-H}
\frac 1 2 \frac {\partial\mathcal{L}} {\partial H} = \lambda_{\mathrm{rel}} \sum\nolimits_{T_{u,v} \neq 0} C_{u,v}(P_u^{\mathrm T}H P_v - S_{u,v})P_u P_v^{\mathrm{T}} + \lambda H .
\end{dmath}

The parameter $\kappa$ controls the peakiness between $Q$ and $\theta$ as shown in Eq.\ref{eq:tran}. The differentiation of $\mathcal{L}$ over $\kappa$ is followed by the chain rule: $\mathcal{L}$ over $\theta$, and then $\theta$ over $\kappa$.
\begin{equation}
\label{eq:grad-kappa}
\frac {\partial\mathcal{L}} {\partial \kappa} = - \lambda_{\mathrm{rev}} \sum\nolimits_{i,f} Q_{if} \Big(M_{if} - \frac{m_i}{z_i} \exp{(\kappa Q_{if}})\Big).
\end{equation}

Note that $\phi_f$ is a stochastic vector, so we optimize the corresponding unnormalized vector $\psi_f$ and then get $\phi_{fw} = \exp{(\psi_{fw})}/z_f$.
The unnormalized word distributions exist in the likelihood term of the reviews corpus.
\begin{equation}
\label{eq:grad-psi}
\frac {\partial\mathcal{L}} {\partial \psi_{fw}} = - \lambda_{\mathrm{rev}} \Big(N_{fw} - \frac {n_f} {z_f} \exp{(\psi_{fw}})\Big).
\end{equation}


For MR3++, the gradients of the $Y$ and $Q$ are given below, and others are the same.

The gradients of the added implicit feature parameters $Y_j$ are computed by
\begin{equation}
\label{eq:grad-Y}
\frac 1 2 \frac {\partial\mathcal{L}} {\partial Y_j} = \sum_{R_{u,i} \neq 0} W_{u,i} |N_u|^{- \frac 1 2} (\hat R_{u,i}^* - R_{u,i}) Q_i + \lambda Y_j, \quad \forall j \in N_u.
\end{equation}

The gradients of the original parameters $Q_i$ are now computed by (an extra term with respect to implicit ratings $Y$ is added to $P$, compared with that in Eq.(\ref{eq:grad-Q}))
\begin{dmath}
\label{eq:grad-Q*}
\frac {\partial\mathcal{L}} {\partial Q_i} = 2 \sum\nolimits_{u:R_{u,i} \neq 0} W_{u,i}(\hat R_{u,i}^* - R_{u,i}) \Big(P_u + |N_u|^{- \frac 1 2} \sum\nolimits_{j \in N_u} Y_j \Big)\\
 - \lambda_{\mathrm{rev}} \kappa \Big(M_i - \frac {m_i} {z_i} \exp{(\kappa Q_i})\Big) + 2 \lambda Q_i.
\end{dmath}

\section{Experiments}\label{paper:Exp}

In the above Section~\ref{paper:m3r} and Section~\ref{paper:m3r-implicit}, we have introduced the solutions to Problem 1 and to Problem 2 respectively. The solution to Problem 1 is the result of the proposed model MR3 (See Eq.(\ref{eq:m3r})), which exploits all three types of information simultaneously. The solution to Problem 2 is the model MR3++ (see Eq.(\ref{eq:m3r-ra})), which extends the MR3 model by incorporating implicit feedback from ratings.

In this section, we first compare our proposed eSMF model with a state-of-the-art Social MF method to show the benefit of exploiting the graph structure of neighbors via trust values which capture the social influence locality. Second, we demonstrate the effectiveness of the proposed model MR3 and the improvement of its extension MR3++ over various different recommender approaches. Further, we design experiments to see the contribution of each data source to the proposed model and the impact of the implicit feedback, followed by sensitivity analysis of our models to three meta parameters.

\subsection{Dataset and Metric}
In this section, we first introduce the two datasets used to evaluate the recommendation performance, including the simple preprocessing and basic statistics. And then we describe the performance metric and the evaluation protocol.

\subsubsection{Datasets and Statistics}{\label{paper:Exp-data}}

\begin{table}[h]
\centering
\tbl{{Statistics of the Datasets}\label{table:data}}{
\begin{tabular}{l c c | c}
\cline{2-3}
 \Xhline{2\arrayrulewidth}
    Statistics              & Epinions  & Ciao       & Total\\
 \hline \hline
 \# of Users                & 49,454    & 7,340      & 56,794\\
 \# of Items                & 74,154    & 22,472     & 96,626\\
 \# of Ratings/Reviews      & 790,940   & 183,974    & 974,914\\
 \# of Social Relations     & 434,680   & 112,942    & 547,622\\
 \# of Words                & 2,246,837 & 28,874,000 & 31,120,837\\
 \hline
 Rating Density             & 0.00022   & 0.0011     & -\\
 Social Density             & 0.00018   & 0.0021     & -\\
 Average Words Per Item     & 30.3      & 1284.9     & -\\
 Average Relations Per User & 8.78      & 15.38      & -\\
 \Xhline{2\arrayrulewidth}
\end{tabular}}
\end{table}

We evaluate our models on two datasets: Epinions and Ciao.\footnote{ \url{http://www.public.asu.edu/~jtang20/} } They are both knowledge sharing and review sites, in which users can rate items, connect to others, and give reviews on products (see Figure~\ref{fig:CiaoExample}). We remove stop words\footnote{\url{http://www.ranks.nl/stopwords}} and then select top $L$ = 8000 frequent words as vocabulary; to avoid noise we reserve users who rated more than three times and remove items that were rated by only once or twice. The items indexed in rating matrix are aligned to documents in doc-term matrix, that is, we aggregate all reviews of a particular item as a `doc'. Statistics of datasets are given in Table~\ref{table:data}. We see that the rating matrices of both datasets are very sparse, and the average length of documents is short on Epinions.



Note that the number of average words per item on Ciao is 42 times longer than that on Epinions, and the social density on Ciao is 12 times denser than that on Epinions. So Ciao contains richer and higher quality information in social relations and reviews. Hence we expect that social relations and reviews contribute much more to the proposed model on the Ciao dataset. We will see the quantitative results later (Section~\ref{paper:more-vs-deep}) which are consistent with this observation.

\subsubsection{Evaluation Protocol and Metric}

There are three kinds of parameters to be set. For optimization-related parameters, we use mini-batch gradient descent method with momentum to optimize the objective functions. Reference to empirical rules (e.g.~\cite{PMF}), we set momentum = 0.8, batchSize = \#training / \#numOfBatches, and learning rate = 0.0007; we randomly shuffle the training data prior to each epoch. For regularization-related parameters, we use norm regularization to avoid over-fitting. According to related bibliographies (e.g.~\cite{TrustSVD}), we set norm penalty $\lambda = 0.5$. For models-related parameters, these are $F$ that is the number of latent factors, $\lambda_{\mathrm{rev}}$ which controls the contribution from reviews, and $\lambda_{\mathrm{rel}}$ which controls the contribution from social relations; see Section~\ref{paper:sensitivity} in detail.

 We randomly select $x$\% as the training set and report the prediction performance on the remaining 1 - $x$\% testing set, where $x$ usually varies in \{20, 50, 80, 90\} percent. The reported results are the average values over five times independent random selection, being similar to 5-fold cross validation. All comparing methods abide by the same evaluation protocol.

The metric {\em root-mean-square error} (RMSE) for rating prediction task is defined as
\begin{equation}
RMSE_{\mathcal{T}} = \sqrt { {\sum\nolimits_{(u,i) \in \mathcal{T}} (R_{u,i} - \hat R_{u,i})^2} \Big{/} {|\mathcal{T}|} },
\end{equation}
where $\mathcal{T}$ is the test set.
Compared with the metric {\em mean absolute error} (MAE)~\cite{CFAlgo99}, RMSE puts more emphasis on large deviation than MAE~\cite{evaluate04}. For instance, an error deviation of 2 points increases the total sum of error by 4 under the metric RMSE, while by 2 under the metric MAE. A smaller RMSE or MAE means a better prediction performance. A small improvement regarding RMSE could have a significant impact on the quality of recommendation and ten percent RMSE improvement led to \$1M Grand Prize~\cite{netflix07,SVDPP}.

There are some other metrics used for evaluation in recommendation systems. For example, in real-world systems we may care about a few top-items for each user and hence we use the metric recall@$K$ to evaluate the performance of recommending the top $K$ items to the target user. In this paper,
We think RMSE is a better metric for our task of rating prediction. In rating prediction, we not only care about the positive preference of ratings above 4 which means the users like the item so much, but also care about the negative preference of ratings under 1 which means that the users do not like the item so much. However, recall@$K$ only care about the positive preference for real recommendation. We focus on the RMSE results and also report the recall results briefly in Section~\ref{paper:recall}.

For each user's ratings, we randomly reserve one into the validation set, and randomly select $1$ or $5$  into the training set, and put the rest into the test set. The former setting is called sparse setting and the latter is dense. We report the test result where the corresponding validation result is the best.

Note that, for the MR3++ model extended from the proposed model MR3, the predicted ratings $\hat R_{u,i}$ (see Eq.(\ref{eq:pred})) should be replaced by $\hat R_{u,i}^*$ accordingly (see Eq.(\ref{eq:svdpp})).

\subsection{Comparing Social MF Methods}{\label{paper:Exp-eSMF}}

\begin{figure}[h]
\centering
\subfigure{ \includegraphics[height=3.8cm,width=1.6in]{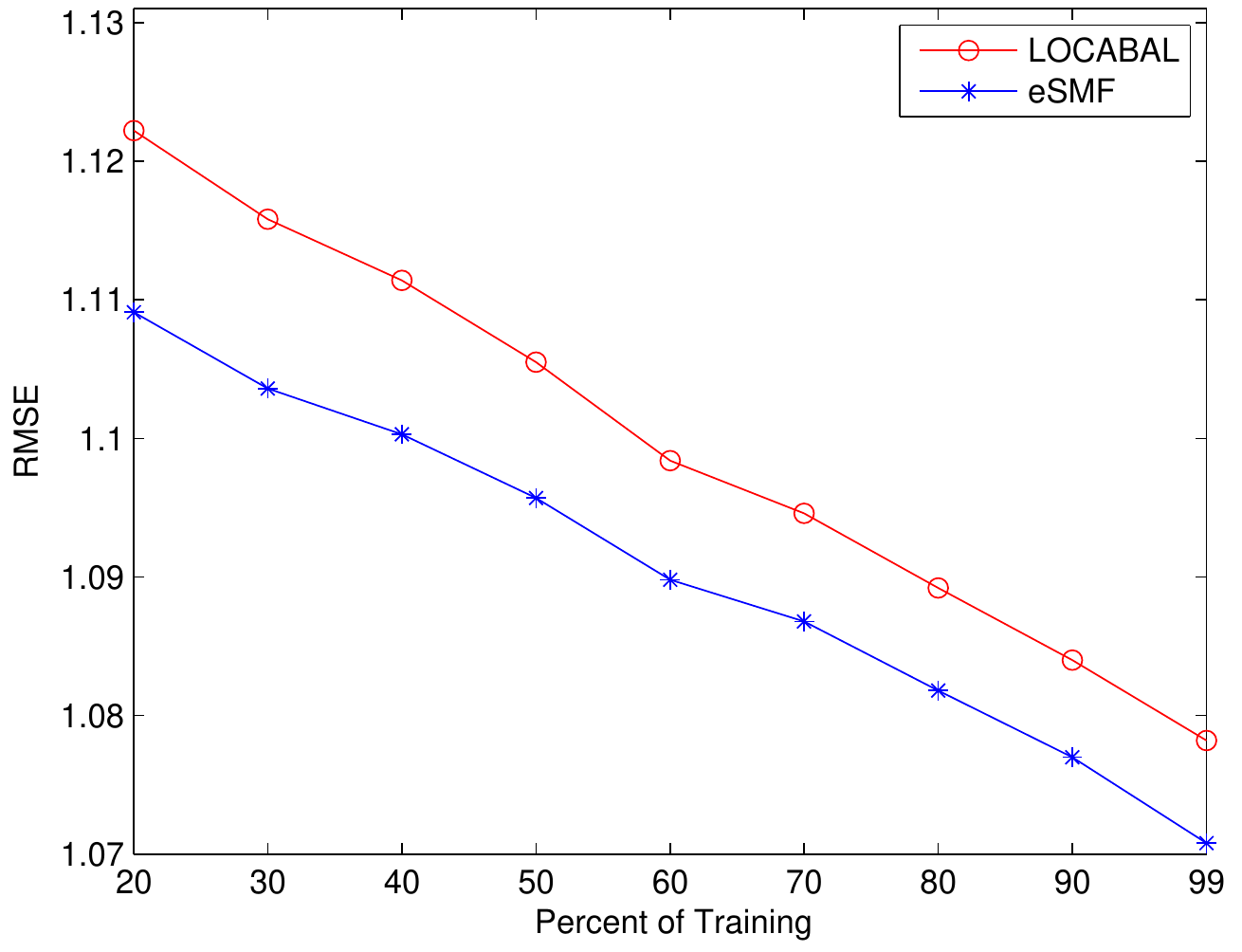} }
\subfigure{ \includegraphics[height=3.8cm,width=1.6in]{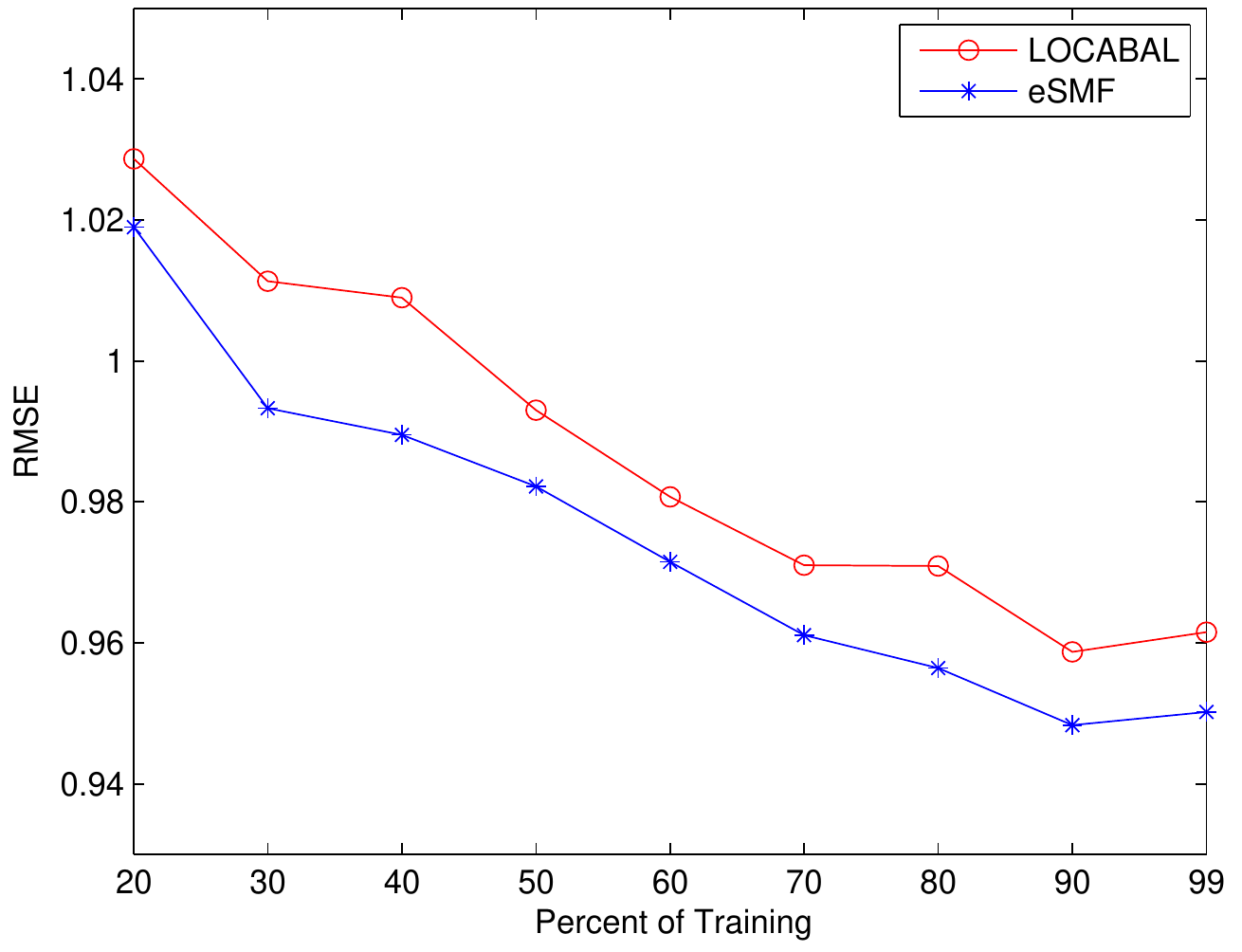} }
\caption{ {\em Comparisons of eSMF with a Social MF method on two datasets.} Left: Epinions; Right: Ciao. The figures are copied from~\cite{MR3}.}
\label{fig:eSMF}
\end{figure}

In this section, we first compare our proposed eSMF model with LOCABAL~\cite{LOCABAL}, a state-of-the-art Social MF method, to show the benefit of exploiting the graph structure of neighbors. The motivation for the comparison is two-fold: 1) to demonstrate that exploiting ratings and social relations more tightly can further improve the performance of social RSs; 2) to form an effective component of the proposed model MR3, which we evaluate in the following section (Section~\ref{paper:Exp-M3R}).

We use grid search to determine $\lambda_{\mathrm{rel}}$ which controls the contribution from social relations and report the best RMSE on the testing set over 50 passes. The eventually reported results are the average values over five times independent random selection of training set, being similar to five-fold cross validation. For both LOCABAL and eSMF, we get the best RMSE when $\lambda_{\mathrm{rel}} = 0.1$. Parameters $\Theta= \{P,Q,H\}$ are randomly initialized from the normal distribution $\mathcal{N}(0,0.01)$.

The results are demonstrated in Figure~\ref{fig:eSMF}, with varying percentage of the training set = \{20, 30, 40, 50, 60, 70, 80, 90, 99\} and we have the following observation:
\begin{itemize}
\item {Exploiting ratings and social relations tightly (i.e., exploiting the graph structure of neighbors via trust values capturing social influence locality) can further improve recommender performance in terms of RMSE on both datasets. For example, eSMF obtains 1.18\%, 0.89\%, and 0.72\% relative improvement compared with a state-of-the-art Social MF method LOCABAL on Epinions with 20\%, 50\%, and 70\% as the training set respectively under the RMSE metric.}
\end{itemize}

\subsection{Comparing Different Recommender Systems}
In this section, we first demonstrate the effectiveness of the proposed model MR3 with different recommendation approaches to show the benefit of modeling three types of data sources simultaneously. And then we demonstrate the improvement of its extension MR3++ to show the benefit of incorporating implicit feedback from ratings.

\subsubsection{Comparing the Proposed Model or MR3 with Different Recommender Systems}{\label{paper:Exp-M3R}}

\begin{table*}
\tbl{{RMSE Comparisons of the Proposed Model MR3 with Different Methods ($F = 10$)} \label{table:mr3}}{
\begin{tabular}{ccccccc|ccc}
\hline \hline
\multicolumn{1}{c}{\multirow{2}{*}{Datasets}} & \multirow{2}{*}{Training} & \multicolumn{5}{c|}{Methods}   & \multicolumn{3}{c}{Improvement of MR3 vs. } \\ \cline{3-7} \cline{8-10}
\multicolumn{1}{c}{}                          &        & Mean   & PMF    & HFT    & LOCABAL & MR3    & PMF          & HFT          & LOCABAL      \\
\hline \hline
\multirow{4}{*}{Epinions} & 20\% & 1.2265 & 1.2001 & 1.1857 & 1.1222  & 1.1051 & 8.60\% & 7.29\% & 1.55\%\\
                          & 50\% & 1.2239 & 1.1604 & 1.1323 & 1.1055  & 1.0809 & 7.35\% & 4.76\% & 2.28\%\\
                          & 80\% & 1.2225 & 1.1502 & 1.0960 & 1.0892  & 1.0648 & 8.02\% & 2.93\% & 2.29\%\\
                          & 90\% & 1.2187 & 1.1484 & 1.0867 & 1.0840  & 1.0634 & 7.99\% & 2.19\% & 1.94\%\\
\hline \hline
\multirow{4}{*}{Ciao}   & 20\% & 1.1095 & 1.0877 & 1.0439 & 1.0287  & 1.0142 & 7.25\% & 2.93\% & 1.43\%\\
                        & 50\% & 1.0964 & 1.0536 & 1.0379 & 0.9930  & 0.9740 & 8.17\% & 6.56\% & 1.95\%\\
                        & 80\% & 1.0899 & 1.0418 & 0.9958 & 0.9709  & 0.9521 & 9.42\% & 4.59\% & 1.97\%\\
                        & 90\% & 1.0841 & 1.0391 & 0.9644 & 0.9587  & 0.9451 & 9.95\% & 2.04\% & 1.44\%\\
\hline \hline
{Average} & & & & & & & 8.34\% & 4.16\% & 1.86\% \\
\hline \hline
\end{tabular}}
\begin{tabnote}
\Note{This table is copied from~\cite{MR3}.}
\end{tabnote}
\end{table*}

We first compare the proposed model MR3 introduced in Section~\ref{paper:m3r} (see Eq.(\ref{eq:m3r}) or Figure~\ref{fig:depend}) with the following different types of recommendation approaches to show the benefit of modeling three data sources simultaneously:

\textbf{Mean.\quad} This method predicts the rating always using the average, i.e. $\mu$ in Eq.(\ref{eq:pred}), across all training ratings. This is the best constant predictor in terms of RMSE.

\textbf{PMF.\quad} This method performs matrix factorization on rating matrix as shown in Eq.(\ref{eq:rating})~\cite{PMF}. It is the representative of latent factors CF (see Figure~\ref{fig:preliminary}(a)). It only uses the rating source.

\textbf{LOCABAL.\quad} This method is based on matrix factorization and exploits local and global social context as shown in Eq.(\ref{eq:socialMF})~\cite{LOCABAL}. It is the representative of Social MF (see Figure~\ref{fig:preliminary}(b))~\footnote{Since we want to compare the performance of MR3 with the state of the art recommenders, so we compare it with LOCABAL and not with our proposed eSMF in Table~\ref{table:mr3}. Instead, the comparison between eSMF and MR3 is conducted in Section ~\ref{paper:contribution-M3R}.}. It only uses ratings and relations.

\textbf{HFT.\quad} This method combines latent factors in ratings with hidden topics in reviews as shown in Eq.(\ref{eq:topicMF})~\cite{HFT}. It is the representative of Topic MF (see Figure~\ref{fig:preliminary}(c)). It only uses ratings and reviews.

\textbf{HFT+LOCABAL.\quad} This is the simple hybrid method of HFT and LOCABAL, which linear weights their results: $\hat{R}_{u,i} = \alpha \hat{R}_{u,i}^{HFT} + (1 - \alpha) \hat{R}_{u,i}^{LOCABAL}$, where $\alpha \in (0, 1)$.

We use the source code PMF\footnote{\url{http://www.cs.toronto.edu/~rsalakhu/}} and HFT\footnote{\url{http://cseweb.ucsd.edu/~jmcauley/}} provided by their authors. For the hyperparameter $\lambda_{\mathrm{rev}}$ which controls the contribution from item reviews, it is determined by grid search. For HFT, $\lambda_{\mathrm{rev}} = 0.1$; and for MR3, $\lambda_{\mathrm{rel}} = 0.001$ and $\lambda_{\mathrm{rev}} = 0.05$. More details about the sensitivity to meta parameters of MR3 will be discussed later (Section~\ref{paper:mr3-hyper-parameters}).

In our experiments, we leave out the comparison with HFT+LOCABAL. Because the core idea of MR3 is the alignment between latent factors found by LOCABAL and hidden topics found by HFT to form a unified model. Theoretically, MR3 is more elegant than linear combination which is out of our motivations. And practically, we do not intent to show the superior of our MR3 over the linear combination. We just show the unified model MR3 is better than its individual components.

The results of the comparison are summarized in Table~\ref{table:mr3}, with varying percentage of the training set = \{20, 50, 80, 90\} and we have the following observations.
\begin{itemize}
\item {Exploiting social relations and reviews beyond ratings can both significantly improve recommender performance in terms of RMSE on the two datasets. For example, HFT and LOCABAL obtain 4.95\% and 5.60\% relative improvement compared with PMF on Epinions respectively,  with 80\% as the training set.}
\item {Our proposed model MR3 always achieves the best results. Compared with HFT and LOCABAL, MR3 averagely gains 0.0466 and 0.0217 absolute RMSE improvement on Epinions and 0.0392 and 0.0165 on Ciao respectively. The main reason is that MR3 jointly models all three types of information effectively. The contribution from each component of data source to MR3 is discussed in the later subsection (Section~\ref{paper:contribution-M3R}).}
\end{itemize}

\subsubsection{Comparing the Extension Model or MR3++ with Different Recommender Systems}{\label{paper:Exp-M3R-RR}}

We then compare the extension of the proposed model {\mbox MR3++} introduced in Section~\ref{paper:m3r-implicit} (see Eq.(\ref{eq:m3r-ra})) to show the benefit of incorporating implicit feedback from ratings.

\textbf{PMF.\quad} This method is investigated in the above subsection (Section~\ref{paper:Exp-M3R}) and we list it here to see the impact of implicit feedback from ratings by comparing it with the following SVD++ method.

\textbf{SVD++.\quad} This method exploits implicit feedback from ratings besides performing matrix factorization on rating matrix as shown in Eq.(\ref{eq:svdpp})~\cite{SVDPP}. We can know the impact of implicit feedback from ratings by comparing it with the PMF method.

\textbf{MR3.\quad} This method is investigated in the above subsection (Section~\ref{paper:Exp-M3R}) and we list it here to see the performance of its extension more clearly. It doesn't mine the data source more deeply , i.e., without incorporating implicit feedback from ratings.

\textbf{MR3++.\quad} This method extends the proposed model MR3 by incorporating implicit feedback from ratings.

We adopt the source code of SVD++ provided in the LibRec.net Java Recommender System Library. For the hyperparameter $\lambda_{\mathrm{rev}}$ which controls the contribution from item reviews and $\lambda_{\mathrm{rel}}$ which controls the contribution from social relations, they are determined by grid search. For \mbox{MR3++}, $\lambda_{\mathrm{rel}} = 0.001$ and $\lambda_{\mathrm{rev}} = 0.005$. More details about the sensitivity to meta parameters of MR3++ will be discussed later.

The results\footnote{The standard deviations are all less than $10^{-5}$. Although, the results in Table~\ref{table:mr3} and Table~\ref{table:mr3-rr} can be merged, we split the whole results into the present two tables to show the contributions of auxiliary sources (i.e., the MR3 model) and the impact of implicit feedback (i.e., the MR3++ model) more clearly.} of the comparison are summarized in Table~\ref{table:mr3-rr} and we have the following observations.
\begin{itemize}
\item {Integrating three data sources and exploiting implicit feedback can both improve the accuracy of rating prediction in terms of RMSE on two datasets. For example, SVD++ and MR3 obtain 7.94\% and 8.02\% relative improvement compared with PMF on Epinions with 80\% as the training set respectively.}
\item {The extension model MR3++ almost achieves the slightly better results. Compared with SVD++ and MR3, MR3++ averagely gains 5.94\% and 0.31\% relative RMSE improvement on Ciao. The main reason is that MR3++ can jointly models all three types of information and meanwhile can mine the rating source deeply. The contributions from the two components of MR3++ is discussed in the following subsection.}
\end{itemize}

\begin{table}
\tbl{RMSE Comparisons of the Extended Model MR3++ with Different Methods ($F = 10$).
\label{table:mr3-rr}}{
\begin{tabular}{cccccc}
\hline \hline
\multicolumn{1}{c}{\multirow{2}{*}{Datasets}} & \multirow{2}{*}{Training} & \multicolumn{4}{c}{Methods} \\
\cline{3-6}
\multicolumn{1}{c}{} &   & PMF   & SVD++    & MR3    & MR3++   \\
\hline \hline
\multirow{4}{*}{Epinions} & 20\% & 1.2001 & 1.1159 & 1.1051 & 1.1026 \\
                          & 50\% & 1.1604 & 1.0816 & 1.0809 & 1.0785 \\
                          & 80\% & 1.1502 & 1.0655 & 1.0648 & 1.0641 \\
                          & 90\% & 1.1484 & 1.0601* & 1.0634 & 1.0618 \\
\hline \hline
\multirow{4}{*}{Ciao}   & 20\% & 1.0877 & 1.0555 & 1.0142 & 1.0132 \\
                        & 50\% & 1.0536 & 1.0276 & 0.9740 & 0.9711  \\
                        & 80\% & 1.0418 & 1.0139 & 0.9521 & 0.9464 \\
                        & 90\% & 1.0391 & 1.0055 & 0.9451 & 0.9425 \\
\hline \hline
\end{tabular}}
\begin{tabnote}
\Note{The columns of PMF and MR3 are copied from Table~\ref{table:mr3} to show the contributions of auxiliary sources and the impact of implicit feedback more clearly. Refer to Section~\ref{paper:more-vs-deep} for the explanation of the star entry.}
\end{tabnote}
\end{table}

\subsection{Contribution of Data Sources and Impact of Implicit Feedback}
In this section, we first measure the contribution from item reviews and social relations in the proposed model MR3, and then measure the impact of implicit feedback from ratings in the extension MR3++ model.

\subsubsection{Contribution of Data Sources from Reviews and Social Relations}{\label{paper:contribution-M3R}}

\begin{figure}[h]
\centering
\subfigure{ \includegraphics[height=3.8cm,width=1.6in]{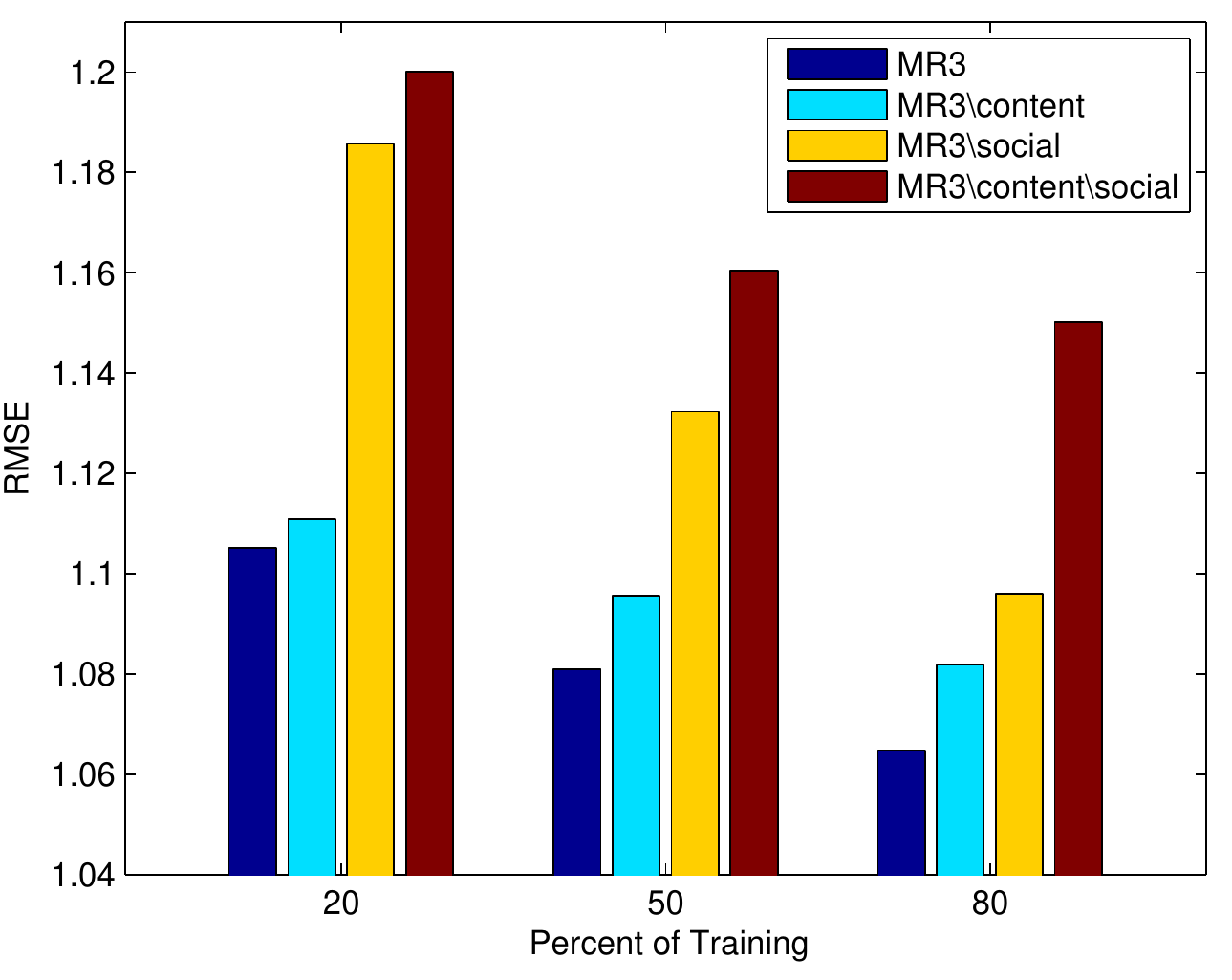} }
\subfigure{ \includegraphics[height=3.8cm,width=1.6in]{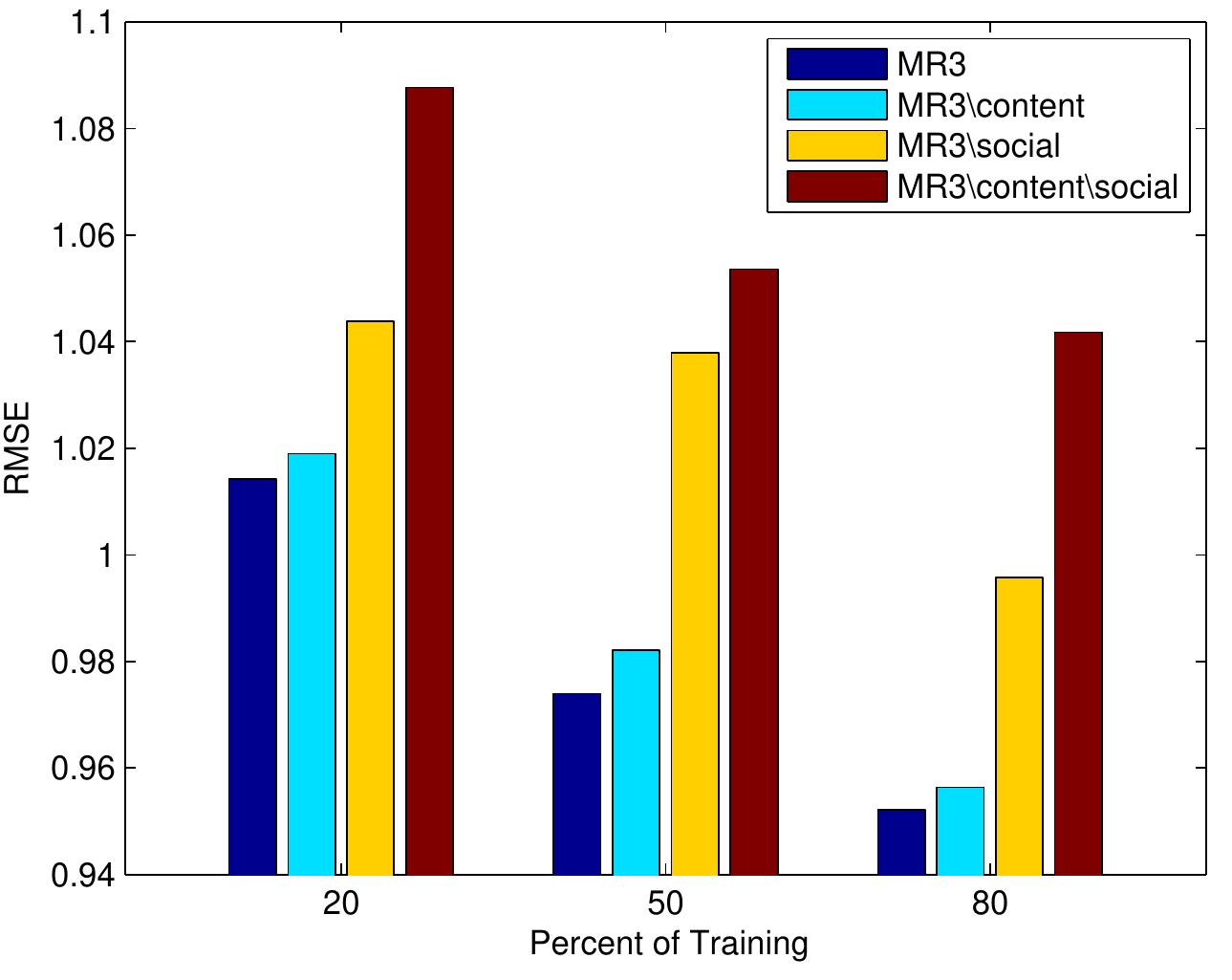} }
\caption{ {\em Predictive performance of MR3 compared with its three components.} Left: Epinions; Right: Ciao. The figures are copied from~\cite{MR3}.}
\label{fig:mr3-component}
\end{figure}

We have shown the effectiveness of integrating ratings with social relations and reviews in our proposed model MR3. We now investigate the contribution of each data source to the proposed model by eliminating the contribution of social relations and reviews from MR3 respectively:

\textbf{MR3$\backslash$content:\quad} Eliminating the impact of reviews by setting $\lambda_{\mathrm{rev}} = 0$ in Eq.(\ref{eq:m3r}), which is equivalent to eSMF as shown in Eq.(\ref{eq:eSMF}).

\textbf{MR3$\backslash$social:\quad} Eliminating the impact of social relations by setting $\lambda_{\mathrm{rel}} = 0$ in Eq.(\ref{eq:m3r}), which is equivalent to HFT as shown in Eq.(\ref{eq:topicMF}).

\textbf{MR3$\backslash$content$\backslash$social:\quad} Eliminating the impact of both reviews and social relations by setting $\lambda_{\mathrm{rev}} = 0$ and $\lambda_{\mathrm{rel}} = 0$ in Eq.(\ref{eq:m3r}), which is equivalent to PMF as shown in Eq.(\ref{eq:rating}).

The predictive results of MR3 and its three components on Epinions dataset are shown in Figure~\ref{fig:mr3-component}. The performance degrades when either social relations or reviews are eliminated. In detail, {\em \mbox{MR3$\backslash$content}}, {\em \mbox {MR3$\backslash$social}}, and {\em \mbox{MR3$\backslash$content$\backslash$social}} averagely reduce 1.19\%, 4.29\%, and 7.99\% relative RMSE performance on Epinions respectively, suggesting that both reviews and social relations contain essential information for recommender.

\subsubsection{Further Analysis of Contribution from Auxiliary Sources} {\label{paper:contribution-further-anlysis}}
We have shown that {\em \mbox{MR3$\backslash$content$\backslash$social}} degrades 7.99\% performance of total relative RMSE; and intuitively we want to know how the additional sources of information help MR3 improve recommendation. We address this issue from the richness perspective of social relations and reviews via contrastive analysis. In detail, we first collect the $(user, item)$ pairs such that the PMF method gets worst prediction while the MR3 method gets the best; we set the difference threshold between PMF and MR3 to 1. For example, given that the prediction error of PMF on $(u,i)$ is $e_1 = |\hat R^{PMF}_{u,i} - R_{u,i}|$ and MR3 is $e_2 = |\hat R^{MR3}_{u,i} - R_{u,i}|$, if $(e_1 - e_2) >= 1$ then we reserve this pair. We then calculate the average social relations among these users and the average review words among these items to measure the quality and richness of the additional information.

With 80\% training on the Epinions dataset, we collect total 4,382 such pairs where the number of users is 3,627 and the number of items is 2,875; they represent 2.80\%, 7.33\%, and 3.88\% on total ratings, total users, and total items, respectively. The average social relations among these users are 14.18 and the average review words among these items are 1129.23. Note that these two measures are much larger than the corresponding total average values which are 8.78 and 30.3 respectively (see Table~\ref{table:data}). These results intuitively confirm the significant contributions from additional data sources to some extent.

\subsubsection{Impact of Implicit Feedback from Ratings}{\label{paper:impact-M3R-RR}}

\begin{figure}[h]
\centering
\subfigure{\includegraphics[height=3.8cm,width=1.7in]{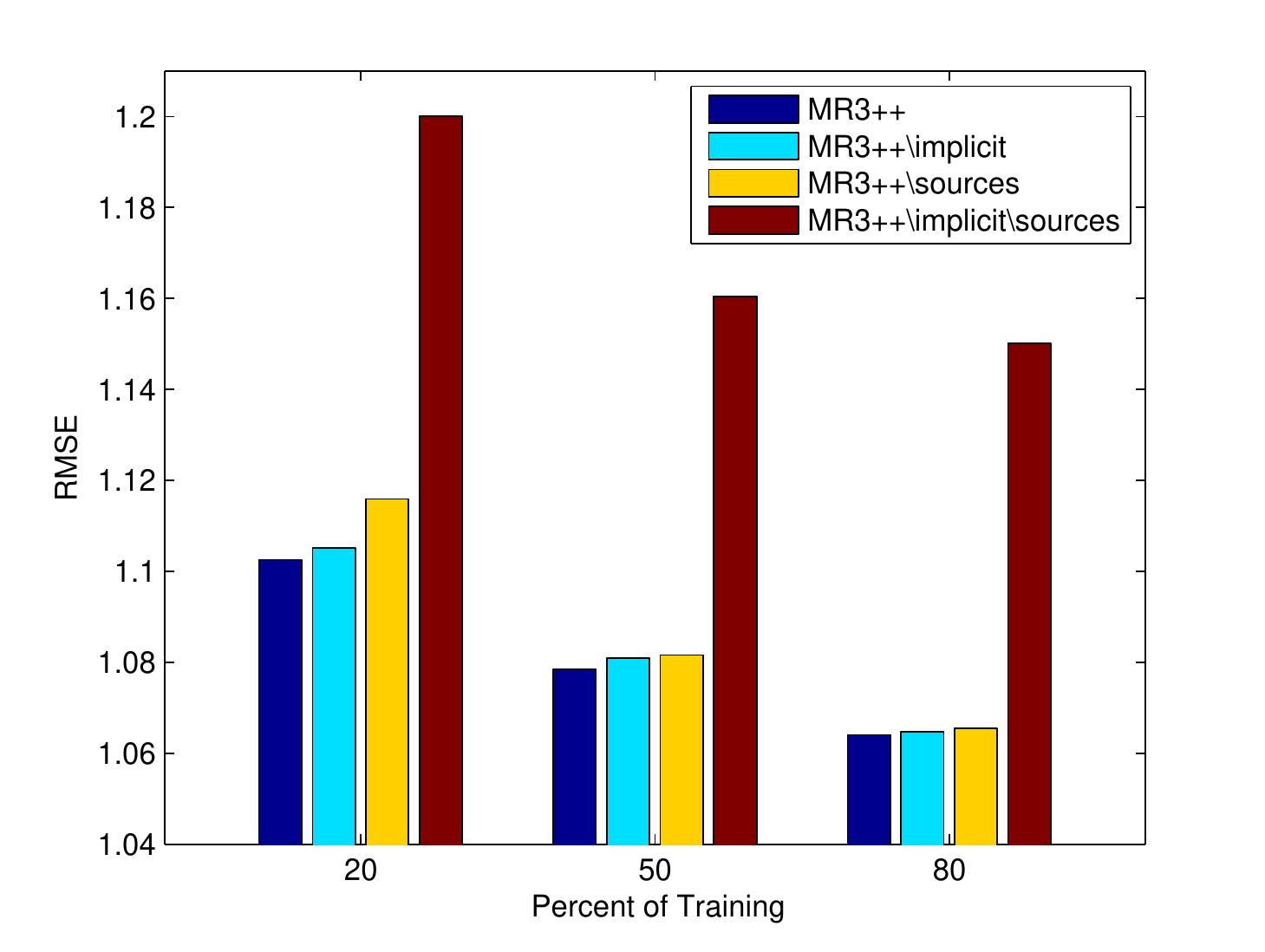}}
\subfigure{\includegraphics[height=3.8cm,width=1.7in]{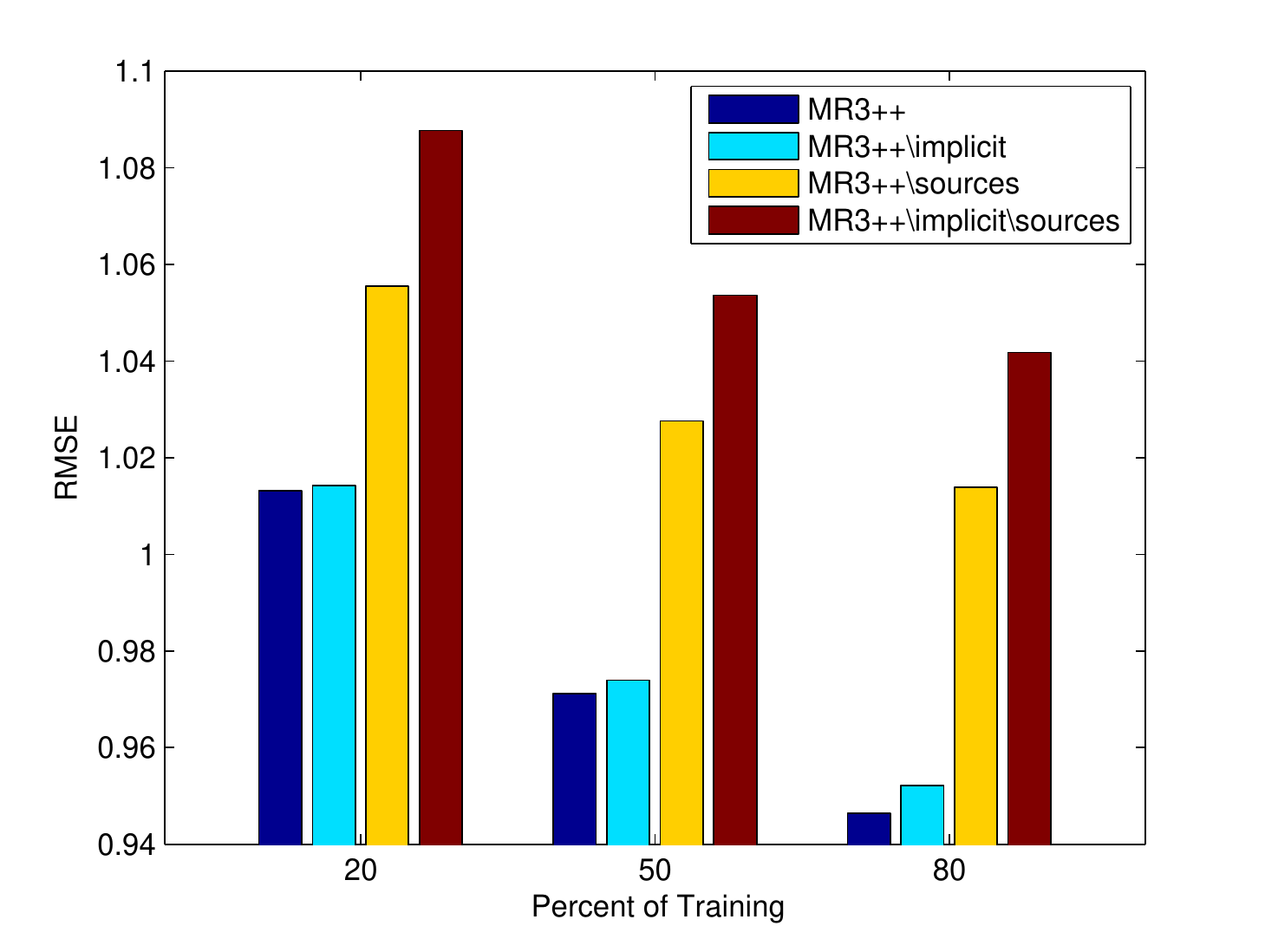}}
\caption{ {\em Predictive performance of MR3++ compared with its two components.} Left: Epinions. Right: Ciao. The figures are copied from~\cite{MR3}.}
\label{fig:mr3pp-component}
\end{figure}

We have investigated the contribution of each data source to the proposed model by eliminating the impact of social relations and reviews from MR3 respectively, showing that extra data sources (item reviews and social relations) are both useful for improving the recommendation performance in terms of RMSE metric. We now investigate the impact of implicit feedback from ratings in the extension model MR3++. The MR3++ model contains two kinds of components: one is to integrate three data sources (ratings, social relations and item reviews) and the other is to mine a single data source deeply (incorporating implicit feedback from ratings). In detail, we investigate the performance of MR3++ by eliminating the impact of data sources and implicit feedback from it in turn:

\textbf{MR3++$\backslash$sources:\quad} Eliminating the impact of more data sources (i.e., remove social relations and item reviews) by setting $\lambda_{\mathrm{rev}} = 0$ and $\lambda_{\mathrm{rel}} = 0$ in Eq.(\ref{eq:m3r-ra}), which is equivalent to the SVD++ model as shown in Eq.(\ref{eq:svdpp}).

\textbf{MR3++$\backslash$implicit:\quad} Eliminating the impact of implicit feedback from ratings by setting $Y_j = 0$ in Eq.(\ref{eq:m3r-ra}), which is equivalent to the MR3 model as shown in Eq.(\ref{eq:m3r}).

\textbf{MR3++$\backslash$sources$\backslash$implicit:\quad} Eliminating the impact of both more data sources and implicit feedback by setting $\lambda_{\mathrm{rev}} = 0$,  $\lambda_{\mathrm{rel}} = 0$ and $Y_j = 0$ in Eq.(\ref{eq:m3r-ra}), which is equivalent to the PMF model as shown in Eq.(\ref{eq:rating}).

The predictive results of MR3++ and its two components on Epinions and Ciao datasets are shown in Figure~\ref{fig:mr3pp-component}. The histograms show that the performance degrades somewhat when implicit feedback are eliminated on both datasets. For example, {\em MR3++$\backslash$implicit} reduces 0.23\% and 0.60\% relative RMSE performance on Epinions and Ciao datasets with 20\% and 80\% as the training set, respectively.

\subsubsection{Integrating more data sources vs. Mining limited data deeply}{\label{paper:more-vs-deep}}
Revisit Table~\ref{table:mr3-rr} and Figure~\ref{fig:mr3pp-component}; we can see clearly that integrating more data sources can improve a bigger margin than only mining a single data source deeply in some cases (here, on the Ciao dataset, in terms of average relative RMSE, the margin greater than 5.61\%), but not always (here, on the Epinions dataset, in terms of average relative RMSE, the margin less than 0.23\%; and when the training percent is 90, i.e. the star entry in Table~\ref{table:mr3-rr}, mining the rating source deeply even outperforms integrating social relation and review data sources).

Consider the quality of social relations and reviews. As we mentioned previously in Section~\ref{paper:Exp-data} (see Table~{\ref{table:data}}), the number of average words per item on Ciao is 42 times longer than that on Epinions, and the social density on Ciao is 12 times denser than that on Epinions. So Ciao contains richer and higher quality information in social relations and reviews. Intuitively and empirically, more data sources may improve performance better than mining single data source deeply when extra data sources contain richer information.

The results on the Epinions dataset in Figure~\ref{fig:mr3pp-component} are worth further thinking. The removing of either implicit feedback component or auxiliary sources component has little effect on the rating prediction performance, but the removing of both components has big impact. This observation is different from the results in Figure~\ref{fig:mr3-component} and even from the results on the Ciao dataset in Figure~\ref{fig:mr3pp-component}. One possible explanation for these findings is that the contributions from the two component are not linear-additive. Different datasets show diverse effect between implicit feedback and auxiliary sources.


\subsection{Sensitivity to Meta Parameters: $F$, $\lambda_{\mathrm{rel}}$ and $\lambda_{\mathrm{rev}}$}{\label{paper:sensitivity}}
We analyze the sensitivity of the proposed model and its extension model to the three important hyperparameters: one controls the contribution from social relations, one controls the contribution from item reviews, and another determines the dimensionality of latent representations.

\subsubsection{Sensitivity Analysis of the Proposed Model or MR3}{\label{paper:mr3-hyper-parameters}}

\begin{figure}[h]
\centering
\includegraphics[height=4.6cm,width=2.4in]{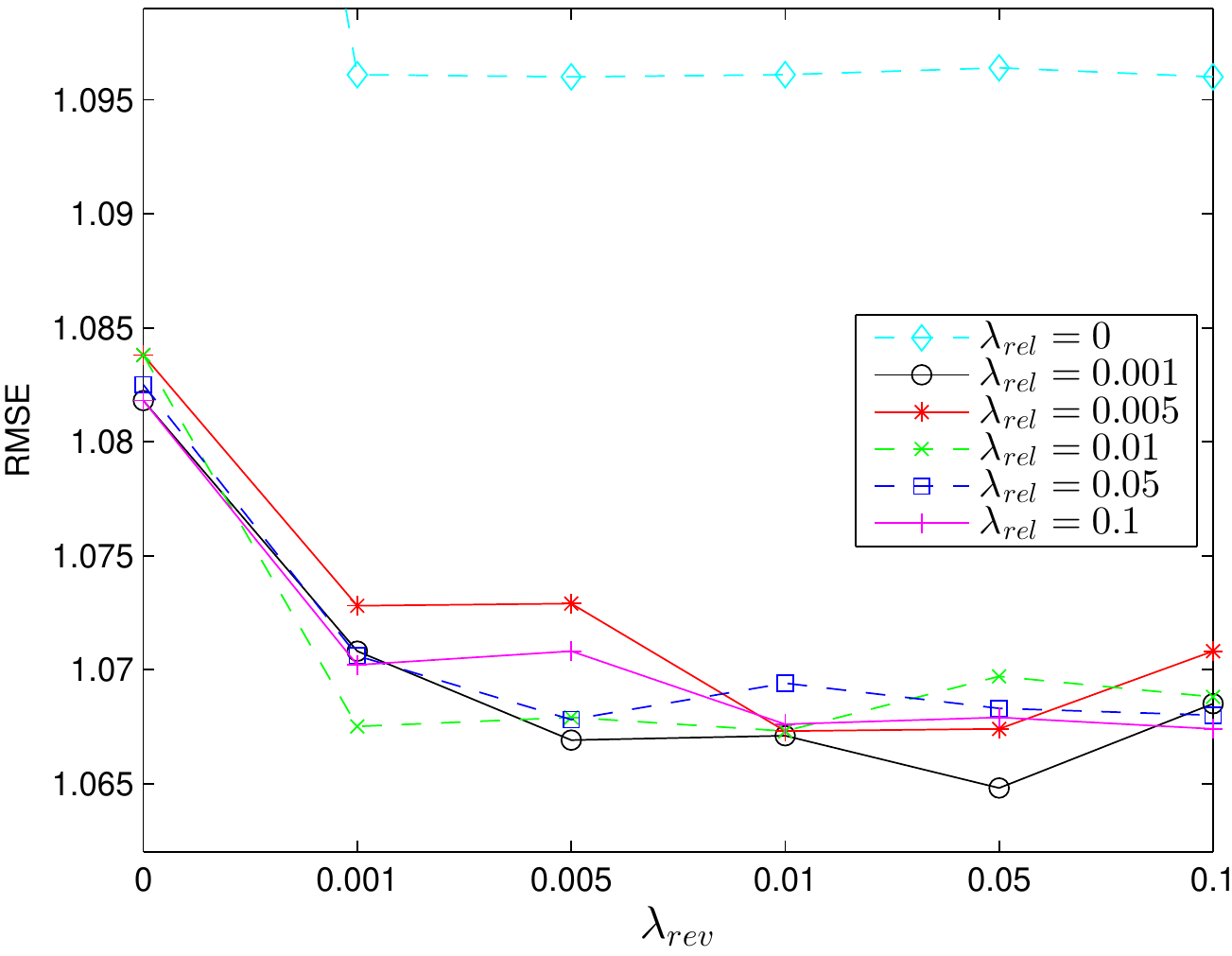}
\caption{ {\em Predictive performance of MR3 by varying $\lambda_{\mathrm{rel}}$ and $\lambda_{\mathrm{rev}}$.} Both vary in \{0, 0.001, 0.005, 0.01, 0.05, 0.1\}. RMSE is 1.1502 when both are zero. Fixing $F$ = 10. Percent of training set = 80. Dataset: Epinions. The figure is copied from~\cite{MR3}.}
\label{fig:mr3relu}
\end{figure}

\begin{figure}[h]
\centering
\subfigure{\includegraphics[height=3.8cm,width=1.6in]{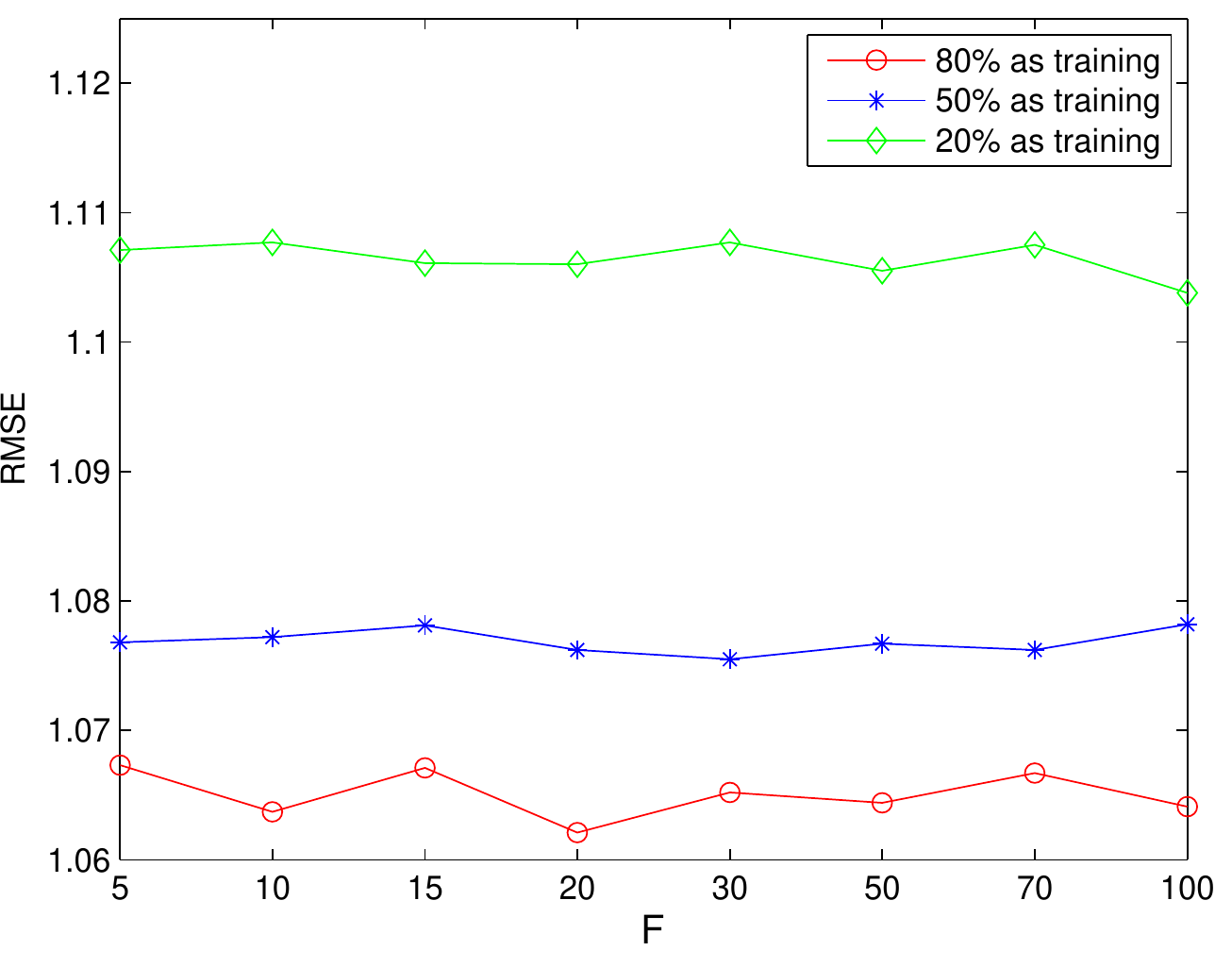}}
\subfigure{\includegraphics[height=3.8cm,width=1.6in]{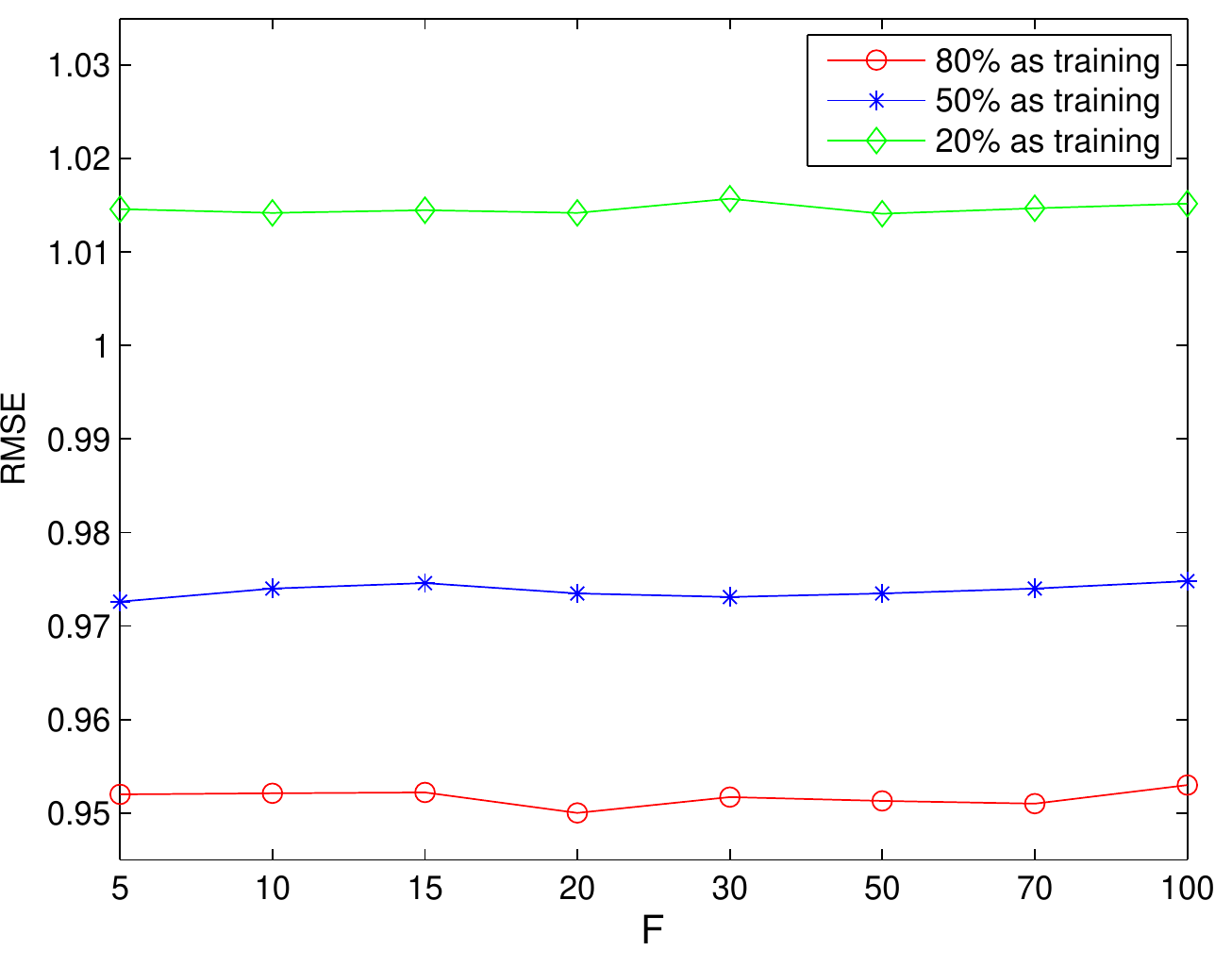}}
\caption{ {\em Predictive performance of MR3 by varying the number of latent factors $F$.} Fixing $\lambda_{\mathrm{rel}} = 0.001$ and $\lambda_{\mathrm{rev}} = 0.05$. Left: Epinions; Right: Ciao. The figures are copied from~\cite{MR3}.}
\label{fig:mr3k}
\end{figure}

The model MR3 has three important parameters: 1) the number of latent factors $F$; 2) the hyperparameter $\lambda_{\mathrm{rev}}$ that controls the contribution from reviews; and 3) the hyperparameter $\lambda_{\mathrm{rel}}$ that controls the contribution from social relations. We investigate the sensitivity of MR3 to these parameters by varying one of them while fixing the other two.

First, we fix $F = 10$ and study how the reviews associated hyperparameter $\lambda_{\mathrm{rev}}$ and the social relations associated one $\lambda_{\mathrm{rel}}$ affect the whole performance of MR3.  As shown in Figure~\ref{fig:mr3relu}, we have some observations: 1) the prediction performance degrades when either $\lambda_{\mathrm{rel}} = 0$ or $\lambda_{\mathrm{rev}} = 0$; (RMSE is 1.1502 when both are zero.) 2) MR3 is relatively stable and not sensitive to $\lambda_{\mathrm{rel}} $ and $\lambda_{\mathrm{rev}}$ when they are small (e.g., from 0.0001 to 0.1), so we choose the reasonable values 0.001 and 0.05 for them respectively.

Next, we fix $\lambda_{\mathrm{rel}} = 0.001$ and $\lambda_{\mathrm{rev}} = 0.05$, and vary the number of latent factors $F = \{5, 10, 15, 20, 30, 50, 70, 100\}$ with 20\%, 50\%, 80\% as the training set respectively. As shown in Figure~\ref{fig:mr3k}, MR3 is relatively stable and not sensitive to $F$, so we choose the reasonable value 10 as default.

\subsubsection{Sensitivity Analysis of the Extension Model or MR3++}{\label{paper:mr3pp-hyper-parameters}}

\begin{figure}[h]
\centering
\subfigure{\includegraphics[height=3.8cm,width=1.6in]{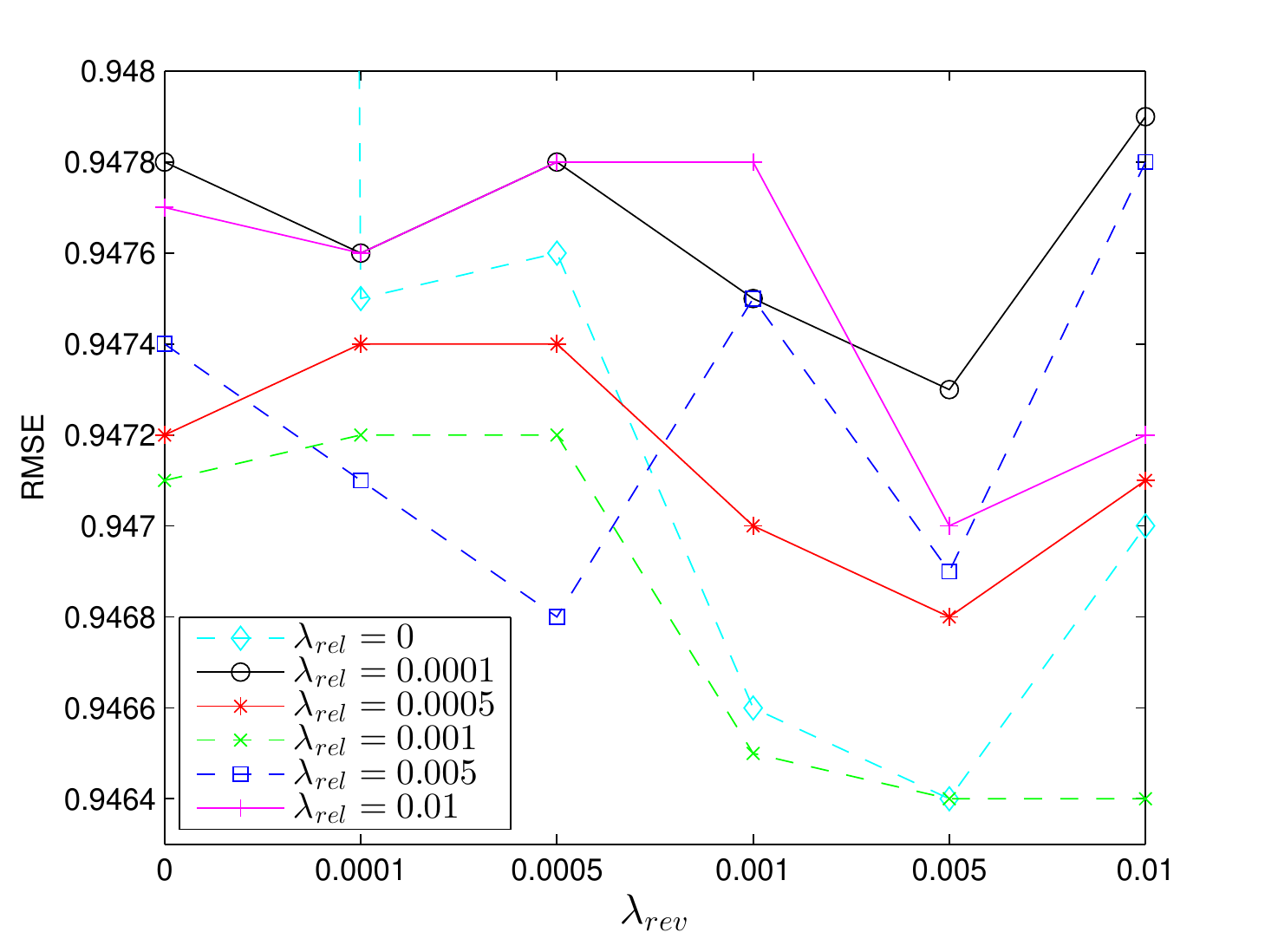}}
\subfigure{\includegraphics[height=3.8cm,width=1.6in]{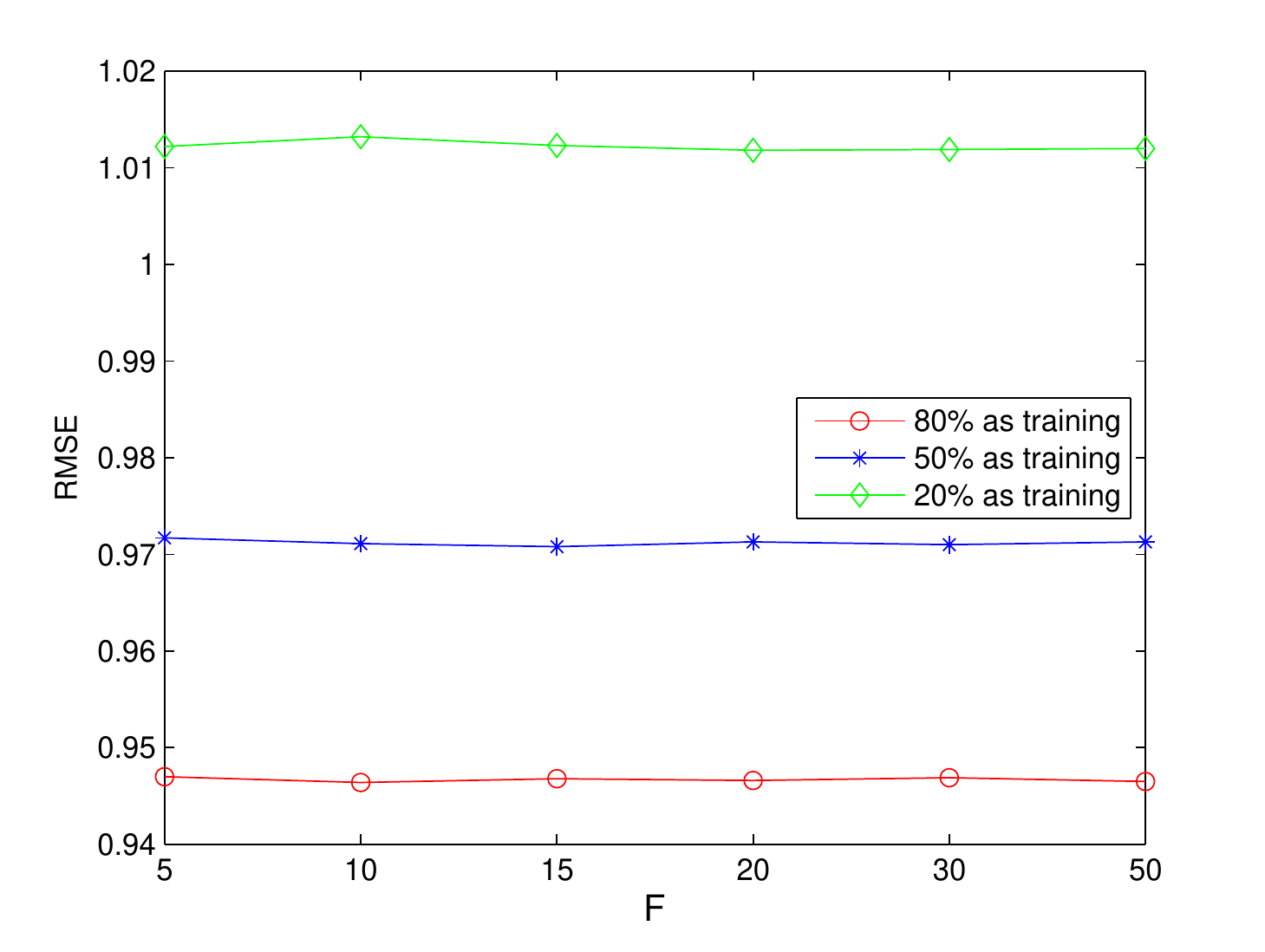}}
\caption{ {\em Predictive performance of MR3++ by varying hyperparameters.}  Left: varying $\lambda_{\mathrm{rel}}$ and $\lambda_{\mathrm{rev}}$ by fixing $F = 10$; RMSE is 1.0139 when both are zero; Percent of Training is 80\%. Right: varying $F$ by fixing $\lambda_{\mathrm{rel}} = 0.001$ and $\lambda_{\mathrm{rev}} = 0.005$. Dataset: Ciao. }
\label{fig:mr3++-f-regus}
\end{figure}

We repeat the above process to investigate the sensitivity of MR3++ to these three meta parameters.

First, we fix $F = 10$ and study how the reviews associated parameter $\lambda_{\mathrm{rev}}$ and the social relations associated one $\lambda_{\mathrm{rel}}$ affect the whole performance of MR3++.  As shown in Figure~\ref{fig:mr3++-f-regus} (on the right), we have some observations: 1) the prediction performance degrades when either $\lambda_{\mathrm{rel}} = 0$ or $\lambda_{\mathrm{rev}} = 0$; (RMSE is 1.0139 when both are zero.) 2) MR3++ is relatively stable and not sensitive to $\lambda_{\mathrm{rel}} $ and $\lambda_{\mathrm{rev}}$ when they are small (e.g., from 0.0001 to 0.01), so we choose the reasonable values 0.001 and 0.005 for them respectively. Note that the length of RMSE range interval is within 0.002.

Next, we fix $\lambda_{\mathrm{rel}} = 0.001$ and $\lambda_{\mathrm{rev}} = 0.005$, and vary the number of latent factors $F = \{5, 10, 15, 20, 30, 50\}$ with 20\%, 50\%, 80\% as the training set respectively. As shown in Figure~\ref{fig:mr3++-f-regus} (on the left), MR3++ is relatively stable and not sensitive to $F$, so we choose the reasonable value 10 as default.

\subsection{Comparing Different Methods with Recall Metric}{\label{paper:recall}}
In this paper, we choose RMSE as the main metric for evaluation because we focus on the rating prediction. However, recall is a more practical metric for real top-$N$ recommender system. Recall cares more about the positive preferences.

 Inspired by top-$N$ recommendations~\cite{FISM}, we define the set of all 5 ratings of a specific user $u$ as positive instances, denoted as $G_u$ for the ground truth and $|G_u|$ is the number of his exact 5 ratings. For each user, we select top variant $K_u = |G_u|$ predicted ratings, denoted as $G'_u$ for evaluation. Thus, the recall, or average hit rate metric is defined as
\begin{equation}\label{eq:hitrate}
recall = \frac{\sum_{u}|G_u\cap G'_u|}{\sum_{u}|G_u|}.
\end{equation}
Hit rate demonstrates the performance of recalling the test items in size $|G_u|$ recommendations list of user $u$. Inversely, if we take the ratings of 1 as negative instances, we can see the performance of predicting dislikes of users.

\begin{table}[h]
	\tbl{Recall Comparisons of Model MR3/MR3++ with Different Methods.
		\label{table:recall-comparison}}{
		\begin{tabular}{ccccccc}
			\hline \hline
			\multicolumn{1}{c}{\multirow{2}{*}{Datasets}} & \multirow{2}{*}{Ratings} & \multicolumn{4}{c}{Methods} \\
			\cline{3-6}
			\multicolumn{1}{c}{} &   & HFT & LOCABAL   & MR3  & MR3++   \\
			\hline \hline
			\multirow{2}{*}{Epinions} & 5 & 74.7127\% & 73.9329\% & 74.6411\% & 74.6455\% \\		
			& 1 & 59.8015\% & 59.2960\% & 60.1507\% & 60.1232\% \\
			\hline \hline
			\multirow{2}{*}{Ciao}   & 5 & 73.4361\% & 73.9042\% & 73.8051\% & 74.0804\% \\
			& 1 & 52.3657\% & 53.1330\% & 53.1330\%  & 53.6445\%  \\
			\hline \hline
		\end{tabular}}
\end{table}

The results of the comparison using recall metric is summarized in Table~\ref{table:recall-comparison}. The setting of meta parameters is the same with Section~\ref{paper:Exp-M3R} and Section~\ref{paper:Exp-M3R-RR}. MR3++ achieves a little improvement over HFT and LOCABAL on two datasets, hitting on either rating 5 or rating 1. Overall, MR3 and MR3++ models got the competitive results. We think it is reasonable that our proposed methods did not achieve the significant improvement with the recall metric. HFT, LOCABAL and our methods focus on getting closer predicted ratings, rather than the best ranking of items. In other words, we define the problem as a prediction regression task, leading us naturally to the standard RMSE metric. The recall metric cares more about the exact match of top-$N$ recommendations where the models are investigated to pick out favorite things of users. These models always put more weight on higher ratings in training set, aiming at recalling the top preferred items. MR3 and MR3++, including HFT, are not designed for this. The emphasis is closer and precise predictions all over 1 to 5 ratings instead.


\subsection{Running Time Analysis}
Practically, the running time of models is a great concern in real-world recommender system. We conduct several rounds of experiments with the same settings, that is, the number of latent factors $F$ = 10, training percent = 80\% and testing the remaining 20\%. We set learning rate = 0.0001 for all models and $nEpoch$ = 5 in Algorithm~\ref{alg:mr3}, meaning we sample and update $z_{d,n}$ every 5 epoches of updating $\Theta, \Phi, \kappa$, which is not required in LOCABAL model.
\begin{table}
	\tbl{Empirical Running Time (seconds) Comparison of the Proposed MR3/MR3++ with Different Methods
		\label{table:time}}{
		\begin{tabular}{cccccc}
			\hline \hline
			\multicolumn{1}{c}{\multirow{2}{*}{Datasets}} & \multirow{2}{*}{Steps} & \multicolumn{4}{c}{Methods} \\
			\cline{3-6}
			\multicolumn{1}{c}{} &   & HFT   & LOCABAL   & MR3   & MR3++   \\
			\hline \hline
			\multirow{3}{*}{Epinions} & Average Epoch & 1.4168 &  1.1251 & 1.549368 &  18.714360 \\
            & Total (best valid) & 77.8941 &  1746.1273 & 212.9282 &  2298.6840 \\
            \cline{2-6}
			& Predicting & 0.2483 &  0.2500 & 0. 274319 &  2.117205 \\			
			\hline \hline
			\multirow{3}{*}{Ciao}   & Average Epoch & 0.3490 & 0.2656 & 0.416025 & 1.622635 \\
            & Total (best valid) & 431.0403 &  152.2076 & 601.7270 &  913.3129 \\
            \cline{2-6}
			& Predicting & 0.0469 & 0.0531 & 0.062847  & 0.098469  \\
			\hline \hline
		\end{tabular}}
		\begin{tabnote}
			\Note{We measure the time cost on a PC machine with Intel(R) Core(TM) i5-4590 CPU @ 3.30GHz, 8G RAM, GCC 5.3.0 installed. Time cost of SVD++ is not listed for we adopt the source code from LibRec.net Java Recommender System Library, and the implementation is quite different from HFT, LOCABAL, MR3 and MR3++.}
			\end{tabnote}
	\end{table}
	
We compare the training time and the predicting time, respectively, as shown in Table~\ref{table:time}. We would like to explain the comparing results as follows.
 \begin{itemize}
  \item As for the time of one training epoch, MR3 is close to HFT and LOCABAL, while MR3 combines more sources of information. MR3++ will include many large-scale implicit feedbacks, so it takes much longer to train through one epoch, especially on the dataset of Epinions which has more users and complicated social networks. We also compare the total time cost for training with different methods. We adopt the early stop on the best valid result to avoid overfitting and we count that time as total.  The total time cost of MR3++ is about ten times of that of MR3, because of the around ten times cost for each epoch. The total time of LOCABAL is extremely long as  it converges after many more iterations than other methods especially on the larger Epinions dataset.
  \item As for the predicting time, it is almost the same comparing MR3 with HFT and LOCABAL. MR3++ costs a little more for predicting because of the extra implicit feedbacks included.
 \end{itemize}

 In a word, from Table~\ref{table:time}, we can conclude that the empirical running time of our proposed models is acceptable compared to other methods.
	

\begin{figure}[h]
\centering
\subfigure[\label{fig:ep-rmse-a}]{\includegraphics[height=3.8cm,width=1.6in]{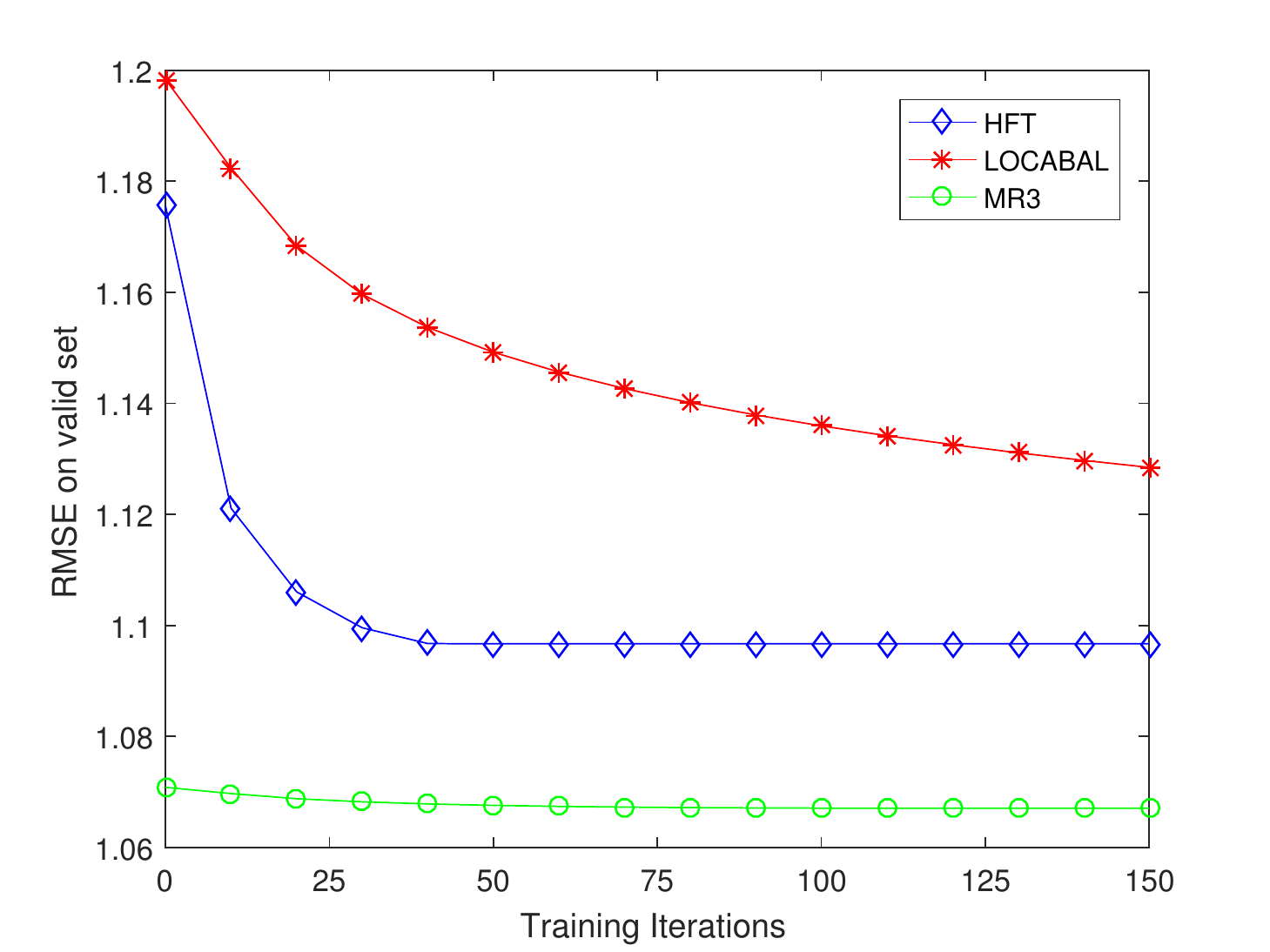}}
\subfigure[\label{fig:ep-rmse-b}]{\includegraphics[height=3.8cm,width=1.6in]{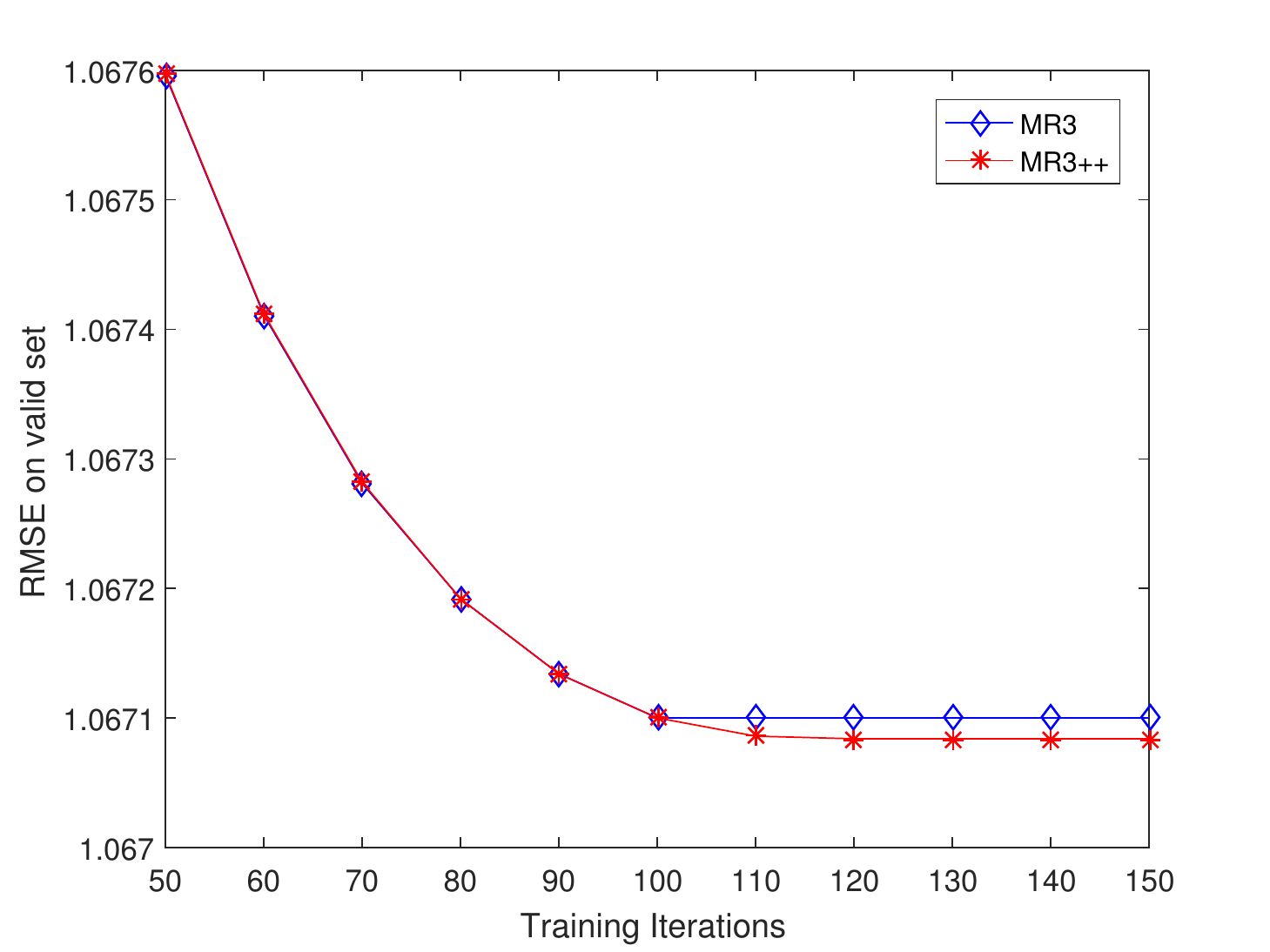}}\\
\subfigure[\label{fig:ciao-rmse-c}]{\includegraphics[height=3.8cm,width=1.6in]{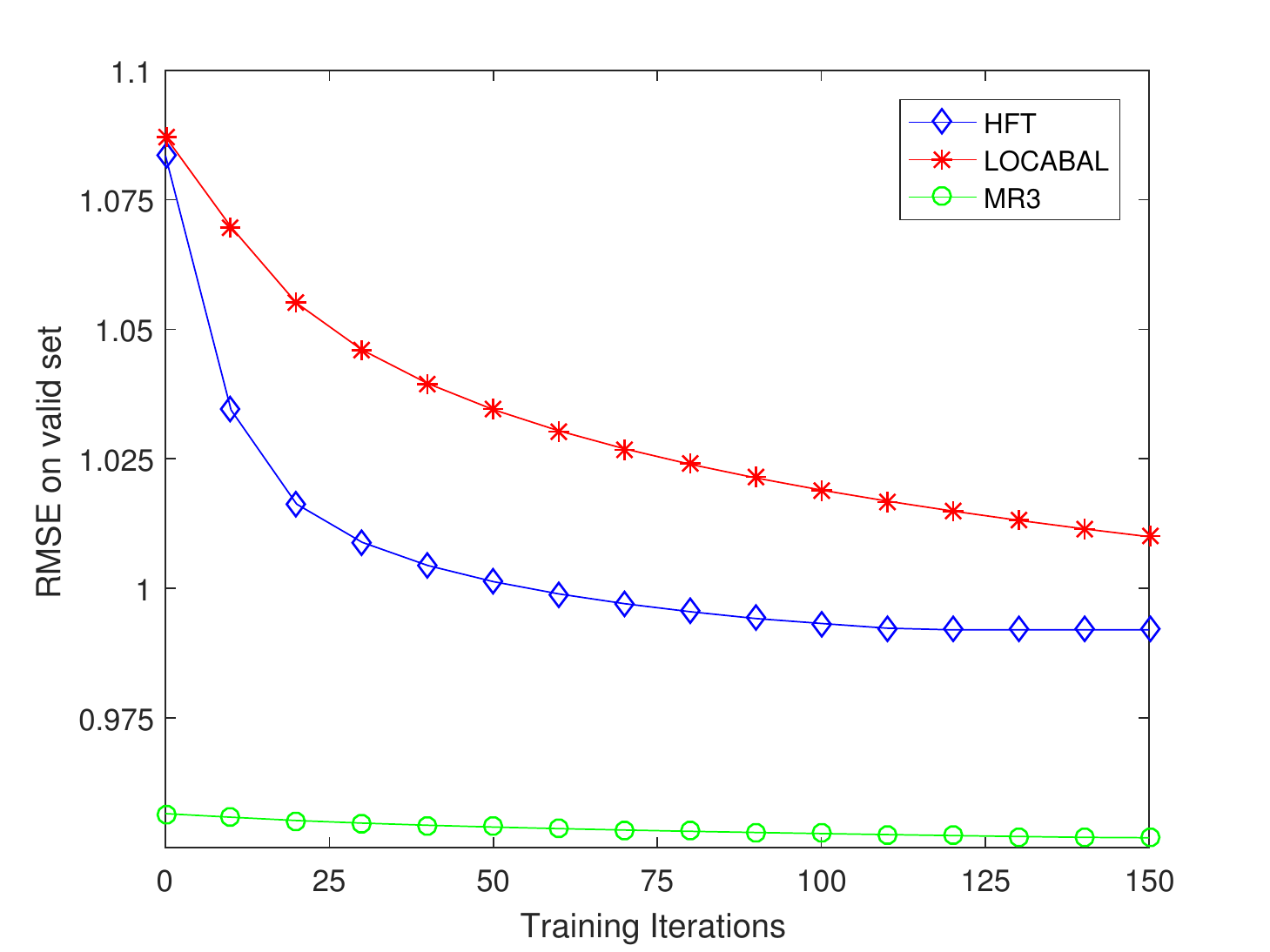}}
\subfigure[\label{fig:ciao-rmse-d}]{\includegraphics[height=3.8cm,width=1.6in]{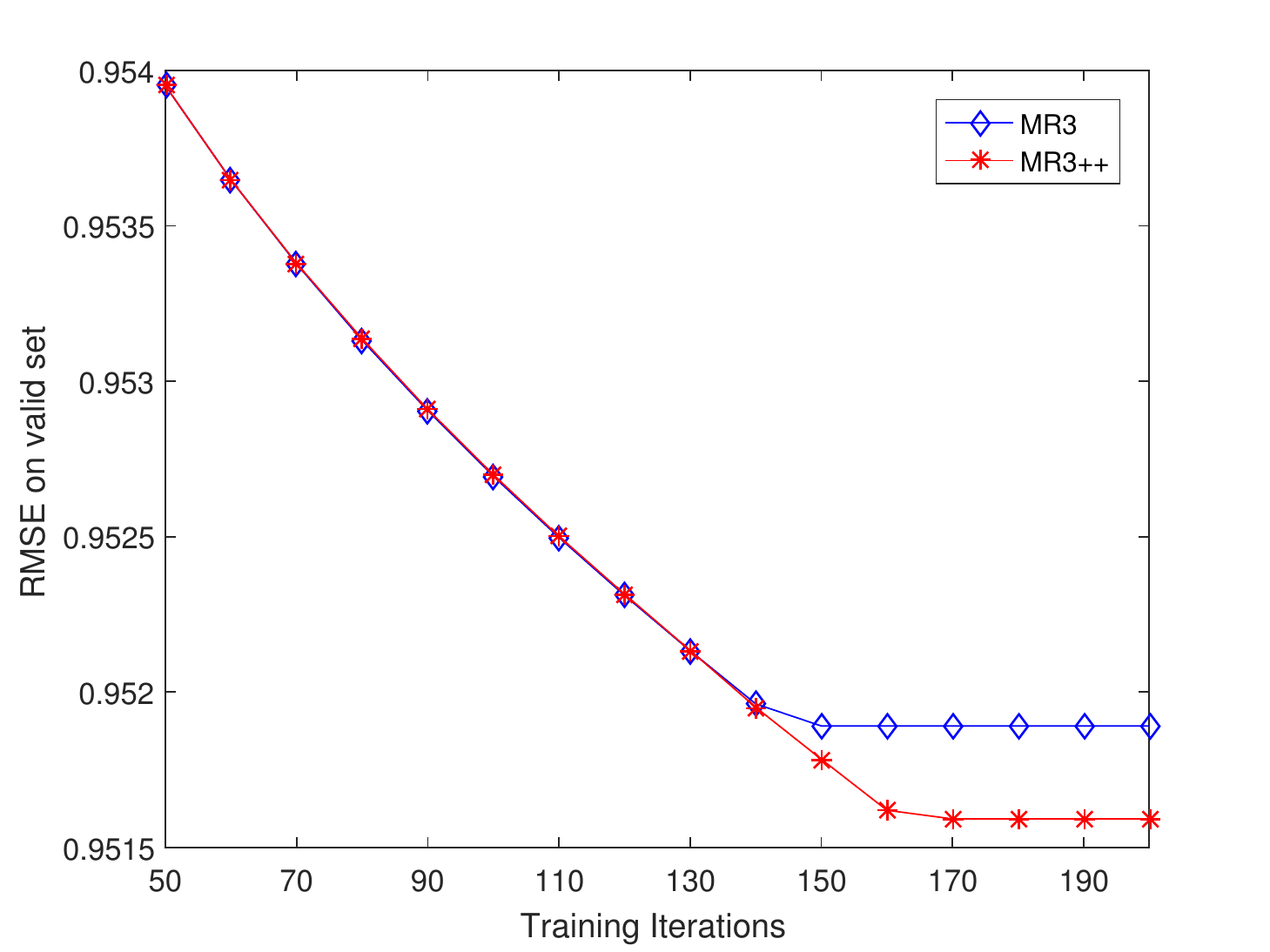}}
\caption{ {\em Training process of all mentioned methods by iterations.} Figure (a) and (b) on dataset Epinions, (c) and (d) on dataset Ciao. (a) and (c) compare HFT, LOCABAL and MR3 on two datasets while (b) and (d) compare MR3 with MR3++ respectively.} \label{fig:iteration-rmse}
\end{figure}

Figure~\ref{fig:iteration-rmse} demonstrates the training process by iterations and RMSE on the valid set. In most cases, these methods become convergent within about 150 rounds of iterations while the valid RMSE of LOCABAL method can still decrease after that. More specifically, about 500 iterations are required on Ciao and even more on Epinions with the conservatively small learning rate. From Figure~\ref{fig:ep-rmse-a} and~\ref{fig:ciao-rmse-c}, MR3 and MR3++ only take few more iterations until convergence yet yield better performance. Interestingly, by Figure~\ref{fig:ep-rmse-b} and~\ref{fig:ciao-rmse-d}, MR3++ continues to converge on a lower valid RMSE value than MR3, for MR3++ takes many implicit feedbacks into account.

\section{Conclusions and Future Work}\label{paper:conclusion}

Heterogenous recommendation information sources beyond explicit ratings including social relations and item reviews present both opportunities and challenges for conventional recommender systems. We investigated how to fuse these three kinds of information tightly and effectively for recommendation by aligning the topic and social latent factors. Furthermore, we mine the limited data source more deeply by incorporating implicit feedback from ratings.

We first proposed a novel model \mbox{MR3} which jointly models ratings, social relations, and item reviews by aligning latent factors and hidden topics to perform social matrix factorization and topic matrix factorization simultaneously for effective rating prediction. Moreover, an extended Social MF method eSMF was obtained by capturing the graph structure of neighbors via trust values to exploit the ratings and social relations more tightly. We then enhanced the proposed model by incorporating implicit feedback from ratings, resulting in the extension model MR3++, to demonstrate its capability and flexibility. The core idea of mining ratings deeply is to learn an extra implicit feature matrix to consider the influence of rated items.

Empirical results on two real-world datasets demonstrated that our proposed models lead to improved predictive performance compared with various different kinds of recommendation approaches. Furthermore, we designed experiments to understand the contribution of each data source and the impact of the implicit feedback from ratings; we also compare the contribution of integrating more data sources with the impact of mining the limited data source deeply from the perspective of the richness of auxiliary information. We finally analyzed the sensitive analysis of our models to the three meta parameters.

We focused on the rating prediction task in this paper and hence we evaluated the performance of recommender systems under the metric RMSE. However, more metrics should be explored in the future work~\cite{accuracyNotEnough}.

\begin{acks}
This work is supported by the NSFC (61472183, 61333014) and the 863 program (2015AA015406). The authors thank Jiliang Tang for providing the datasets. The authors would also like to thank the reviewers for their profound comments and valuable suggestions to improve the quality of this paper.
\end{acks}

\bibliographystyle{ACM-Reference-Format-Journals}
\bibliography{acmsmall-sample-bibfile}


\end{document}